\documentclass[twocolumn]{aastex631}

\def\hb{H$\beta$}

\def\1h07{1H\,0707$-$495}
\usepackage{ulem}

\def\aap{A\&A\/}
\def\apjs{ApJS}

\def\apjl{ApJL}

\def\civ{{C\sc{iv}}\/}
\def\ciii{{C\sc{iii}]}\/}
\def\mgii{{Mg\sc{ii}}\/}
\def\hb{{H$\beta$\/}}

\begin{document}

 \title{Probing the CIV continuum size luminosity relation in active galactic nuclei with photometric reverberation mapping}

\author[0000-0002-5854-7426]{Swayamtrupta Panda}\thanks{CNPq Fellow}\email{spanda@lna.br}
\affiliation{Laborat\'orio Nacional de Astrof\'isica (LNA), Rua dos Estados Unidos 154, Bairro das Na\c c\~oes, CEP 37504-364, Itajub\'a, MG, Brazil}

\author[0000-0002-6716-4179]{Francisco Pozo Nu\~nez}
\affiliation{Astroinformatics, Heidelberg Institute for Theoretical Studies, Schloss-Wolfsbrunnenweg 35, 69118 Heidelberg, Germany}

\author[0000-0002-2931-7824]{Eduardo Ba\~nados}
\affiliation{Max-Planck Institut f\"ur Astronomie, K{\"o}nigstuhl 17 Heidelberg, Germany}

\author[0000-0002-0320-1292]{Jochen Heidt}
\affiliation{Landessternwarte, Zentrum f\"ur Astronomie der Universit\"at Heidelberg, K{\"o}nigstuhl 12, 69117 Heidelberg, Germany}



\begin{abstract}

Reverberation mapping accurately determines virial black hole masses only for redshifts $z <$ 0.2 by utilizing the relationship between the \hb{} broad-line region (BLR) size and the 5100\AA\ continuum luminosity established with $\sim 200$ active galactic nuclei (AGN). For quasars at $z \sim 2-3$ determining the BLR size is time-consuming and limited by seasonal gaps, requiring e.g., $\sim$ 20 years of monitoring of the \civ{} emission lines. In this work, we demonstrate that an efficient alternative is to use a continuum size-luminosity relation, which can be obtained up to 150 times faster than BLR sizes using photometric reverberation mapping (PRM). We outline the method and its feasibility based on simulations and propose an observational strategy that can be carried out with meter-class telescopes. In particular, we focus on the ESO La Silla 2.2 meter telescope as it is suitable for an efficient PRM campaign.
These observations will provide the scaling factor between the accretion disk and the BLR size (for \civ{}-1350\AA), which is crucial for estimating the masses of black holes at higher redshifts ($z \gtrsim 2-3$).

\end{abstract}



\section{Introduction} \label{sec:intro}

Reverberation mapping (RM) can provide accurate virial black hole masses ($M_{\rm BH}$) for redshift ($z$) below 0.2 using the well-known relationship between the size of the broad line region (BLR) and the 5100\AA\ continuum luminosity \citep[$R_{\rm{BLR}} \propto L_{\rm{AGN}}^{\alpha}$,][]{1991ApJ...370L..61K, 2000ApJ...533..631K, 2013ApJ...767..149B, 2019ApJ...886...42D, 2023FrASS..1030103P}. 
Photoionization models, assuming a constant continuum shape and BLR density, originally predicted this correlation with $\alpha = 0.5$ (\citealt{1979RvMP...51..715D}; \citealt{1990agn..conf...57N}).
However, a study by \cite{2000ApJ...533..631K} involving 17 low redshift quasars ($z < 0.3$) suggested $\alpha = 0.7$. 
Subsequent larger RM studies, particularly those correcting for host galaxy contamination and focusing on \hb{} and optical flux, have found $\alpha$ close to 0.5 (e.g., \citealt{2009ApJ...705..199B}; \citealt{2013ApJ...767..149B}).

To date the $R_{\rm BLR}-L_{5100\text{\AA}}$ relation has been established using almost 200 AGN in the \hb{} region\footnote{Recent studies have improved the \textit{R - L} relation also in the \mgii{} region ($z \sim 0.004 - 1.89$) expanding the sample size to $\gtrsim$200 AGN \citep{2023arXiv230916516C, 2023arXiv231003544Z, 2023arXiv230501014S}, although the presence of the Fe{\sc ii} pseudocontinuum underneath and around the \mgii{} emission line complicates the already complex kinematics for this region \citep{2019MNRAS.484.3180P, 2019ApJ...875..133P, 2023arXiv231003544Z, 2024arXiv240118052P}.} (see \citealt{2019FrASS...6...75P}; \citealt{2023arXiv230501014S} and references therein) and its extrapolation allows $M_{\rm BH}$ estimation for high-z ($z\sim6$) quasars (e.g., \citealt{2024arXiv240213319L}), albeit with large uncertainty. 
Attempting to calibrate the $R_{\rm BLR}-L$ relation with quasars at higher $z$ (e.g., $z$$\sim2$ using the \civ{} line) is exceptionally time-consuming, requiring campaigns spanning $\sim$20 years to detect delays (\citealt{2018ApJ...865...56L}; \citealt{2019ApJ...887...38G}; \citealt{2019MNRAS.487.3650H}; \citealt{2021ApJ...915..129K}). 
Seasonal gaps further compromise the accuracy of these delay measurements, necessitating light curve interpolation and modeling, which introduces additional complexities and potential sources of error. 

An efficient alternative involves RM of the continuum emission from the accretion disk (AD). 
The AD continuum size-luminosity relation ($R_{\rm AD}-L$) can be established more rapidly, given the AD's approximately tenfold smaller size compared to the \hb{}-emitting BLR (see \citealt{2023ApJ...948L..23W} and references therein). 

The standard AGN AD theory posits that the effective temperature of a thin disk, varying with radius, depends on black hole mass and accretion rate (\citealt{1969Natur.223..690L}; \citealt{1972A&A....21....1P}; \citealt{1973A&A....24..337S}; \citealt{1973blho.conf..343N}; \citealt{1981ARA&A..19..137P};  \citealt{1987ApJ...321..305C}; 
\citealt{1994ApJ...428L..13N};
\citealt{1998ApJ...500..162C}; \citealt{2005ApJ...622..129S}; \citealt{2007MNRAS.380..669C}; \citealt{2011A&A...525L...8C};
\citealt{2014ARA&A..52..529Y}; \citealt{2023Univ....9..492P}). 
Therefore, the AD's radial extent can be investigated by studying the continuum emission at different wavelengths (see \citealt{2023MNRAS.522.2002P} and references therein). 
Based on the reprocessing AD scenario and due to light travel time effects, the inner regions of the AD, which are detected by shorter wavelengths, react first to irradiation by the so-called X-ray corona, while the outer parts, which are detected by longer wavelengths, react later and with a time delay $\tau_{\rm AD}$ (\citealt{2021iSci...24j2557C}). 
These delays provide valuable information about the size ($R_{\rm AD}\sim c\cdot\tau_{\rm AD}$) and the temperature stratification across the AD. 
They can be measured using photometric reverberation mapping (PRM), which can use a combination of broad, medium, and narrow-band photometry (\citealt{2017PASP..129i4101P}; \citealt{2019NatAs...3..251C}) to track variations of carefully selected emission line-free continuum regions.

RM studies of the AD suggest a delay-wavelength relation $\tau\propto\lambda^{4/3}$, consistent with geometrically thin AD models. 
However, observed AD sizes are several times ($\sim 3$ to 5) larger than anticipated by standard AGN AD theory (\citealt{2014ApJ...788...48S}; \citealt{2015ApJ...806..129E}; \citealt{2016ApJ...821...56F}; \citealt{2018ApJ...857...53C}; \citealt{2023MNRAS.525.4524G}). 
Previous studies utilizing independent microlensing techniques have produced comparable outcomes (e.g., \citealt{2007ApJ...661...19P}; \citealt{2012ApJ...756...52M}; \citealt{2013ApJ...769...53M}; \citealt{2016AN....337..356C}).
Termed the ``\textit{accretion disk size problem}," its implications for the standard disk-reprocessing scenario remain debated.
Dedicated monitoring initiatives utilizing SWIFT and HST telescopes (e.g. NGC5548 of \citealt{2015ApJ...806..129E}) have significantly improved our understanding of AD sizes at different wavelengths. 
The observed AD size between 1350\text{\AA} and 1647\text{\AA} of about 0.10 days for NGC 5548 is significantly larger than the 0.035 days predicted based on $M_{\rm BH}$ and accretion rate, which challenges the theoretical predictions of the alpha disk model. 
However, it is important to note that the time delay measurements are subject to a considerable uncertainty of about 50 percent and that the interpolation step used in the cross-correlation analysis exceeds the resolution of the predicted time delay. 
These factors emphasize the importance of conducting observations with a finer temporal resolution, especially for low-redshift sources, to resolve these discrepancies and to corroborate the theoretical models.

There are proposed solutions to the AD size problem that include contamination from nearby regions, such as the BLR, which appears in the form of lines and diffuse continuum emission (DCE; \citealt{2001ApJ...553..695K}; \citealt{2018MNRAS.481..533L}; \citealt{2019MNRAS.489.5284K}; \citealt{2022MNRAS.509.2637N}; \citealt{2023A&A...680A.102P}) or even other under-appreciated non-disk components (\citealt{2019NatAs...3..251C}). 
These components can lead to lower or higher delays depending on the relative contribution to the filters (\citealt{2023MNRAS.522.2002P}).
Internal reddening due to dust near AD, and more distant host galaxy contamination further complicate luminosity determinations, potentially underestimating AD sizes (\citealt{2023MNRAS.519.4082G}).
Since internal extinction is significant in most AGN, this can lead to nuclear luminosities that are underestimated up to a factor of 4 and 10 in the optical and UV, respectively (\citealt{2017MNRAS.467..226G}). 
In addition, the contamination from the host galaxy is considerable and several efforts have been made to minimize its contribution so that the AGN luminosities are correctly determined (see e.g., \citealt{2022A&A...657A.126G}).
Alternatively, models of X-ray illumination could also explain the observed larger delays for certain cases where the corona is located at a distance of more than $\sim40$ gravitational radii above the black hole (\citealt{2022A&A...666A..11P}), although accounting for scattering due to the BLR can show time delays that are similar to the effect of the rising height of the X-ray source (\citealt{2023A&A...670A.147J}). Simultaneous observations for the BLR and AD have shown that overly massive black holes could also explain the larger observed AD sizes (\citealt{2019MNRAS.490.3936P}). 
This suggests that the unknown geometry of the BLR+AD system may lead to a significant underestimation of the black hole mass by the virial product and, thus, biased AD measurements.

Only 21 local ($z \lesssim 0.2$) objects offer high-quality continuum time delay measurements (\citealt{2023ApJ...948L..23W}). 
While a relation between DCE size and 5100\AA\ luminosity akin to the \hb{} BLR size - $5100$\,\AA\ luminosity relation is noted (\citealt{2022MNRAS.509.2637N}), the detected continuum delays (at 5100\AA) are still a factor of ten shorter than typical BLR time delays ($\tau_{\rm{H}\beta}/\tau_{5100} \sim 10$) and follow $\tau_{5100} \propto L_{5100}^{1/2}$. 
The scaling factor allows us to estimate the size of the BLR and, together with the velocity dispersion of the emission line, to calculate the mass of the black hole. 
However, it is not clear whether this relationship and the scaling factor also apply to more luminous quasars at higher redshifts. 

In this work, we aim to study the feasibility of PRM of selected continuum regions of high-redshift quasars to provide the scaling factor required to estimate the BH mass.

\section{The sample}\label{sec2}

We have selected the sample studied in \cite{2021ApJ...915..129K} (their table 6), which provides high-quality RM measurements of the \civ{} emission line and black hole mass data for 38 AGN across a diverse range of redshift ($0.001 < z < 3.4$) and luminosity (39.9 $<$ log $\lambda L_{\lambda}(1350\text{\AA}) <$ 47.7, in erg s$^{-1}$).
In addition to this compilation,  \cite{2021ApJ...915..129K} demonstrates time-lag recovery from their long-term monitoring efforts which spanned approximately 20 years and focused on the most luminous and highest-redshift quasars of the sample ($2.2 < z < 3.2$) in the northern hemisphere, yielding high-quality light curves crucial for reliable \civ{} lag measurements. 
Their compilation integrates the findings of \cite{2018ApJ...865...56L}, who conducted a similar long-term monitoring campaign of about 10 years for 17 high-luminosity quasars located in the south, contributing BLR \civ{} sizes and $M_{\rm BH}$ for 8 high-redshift ($2.5 < z < 3.4$) quasars. 
These two studies, \cite{2021ApJ...915..129K} and \cite{2018ApJ...865...56L}, collectively represent the most extensive spectrophotometric RM investigations of quasars to date.

\section{Simulations}\label{sec3}

The expected AD time delays are calculated following the standard thermal reprocessing scenario with its application to PRM of the accretion disk, as outlined in \cite{2019MNRAS.490.3936P,2023MNRAS.522.2002P}. 
In brief, the energy flux radiated from the disk combines viscous heating and external irradiation from the so-called X-ray corona. 
The temperature across the disk is derived from these two energy components. It is proportional to $M_{\rm BH}$ and the mass accretion rate ($\dot M$), $T(r)\propto (M_{\rm BH} \dot M)^{1/4}r^{-3/4}$, as described by the standard AGN accretion disk theory (\citealt{1969Natur.223..690L}; \citealt{1972A&A....21....1P}; \citealt{1973A&A....24..337S}; \citealt{1973blho.conf..343N}; \citealt{1998ApJ...500..162C}; \citealt{2005ApJ...622..129S}; \citealt{2007MNRAS.380..669C}; \citealt{2018ApJ...866..115P}).

The observed AD-UV/optical continuum emission, $F_{c}(\lambda, t)$, is obtained from the convolution of the X-ray driving light curve, $F_{x}(t)$, with the transfer function
$\Psi(\tau|\lambda) \propto \partial B_{\nu}(\lambda,T(t-\tau)) / \partial L_{x}(t-\tau)$, so that $F_{c}(\lambda, t) = F_{x}(t) * \Psi(\tau|\lambda)$, where $B_{\nu}$ is the Planck function of a blackbody characterized by the radial temperature profile of the disk $T(t-\tau)$. 
The boundary of the disk is assumed to be $R_{0}\sim6 R_{g}$, where $R_{g} = 2GM_{\rm BH}/c^2$ is the Schwarzschild radius \citep{2002apa..book.....F, 2016ASSL..440....1L, 2019NatAs...3...41R, 2022MNRAS.510.1010P}. 
The X-ray corona is supposedly at the height $h = 10 R_{g}$ \citep{2021ApJ...907...20K, 2021MNRAS.503.4163K, 2023A&A...670A.147J} and drives the emission, which is further reprocessed on the disk. 
The disk emission reaches the observer with a time delay $\tau \propto (r^2+h^2)^{1/2}$, based on the assumption of a Keplerian disk structure.

We model $F_{x}(t)$ using the method of \citet{1995A&A...300..707T}, assuming a power spectral density $P(\nu)\propto \nu^{-\alpha}$, where  $\alpha = 2.0$, which is consistent with the random walk process observed in quasars light curves (\citealt{1999MNRAS.306..637G}; \citealt{2001ApJ...555..775C}; \citealt{2007A&A...462..581H}; \citealt{2017ApJ...834..111C}).
At this stage, our simulations are noise-free, and the light curves have a total duration\footnote{It is recommended to first generate light curves that are significantly longer, ideally about 10 times as long as the observed data set. This strategy effectively addresses the issue of 'red noise leak' (\citealt{2003MNRAS.345.1271V}). In cases where this approach is not feasible, alternatives such as de-trending the light curves can be used. This method helps to reduce the effects of sinusoidal trends whose periods exceed the duration of the light curve (\citealt{2018MNRAS.474.3237L}).} of $T = 1000$ days and an ideal sampling of $\Delta t = 0.1$ days.

\begin{figure}
    \centering
    \includegraphics[width=\columnwidth]{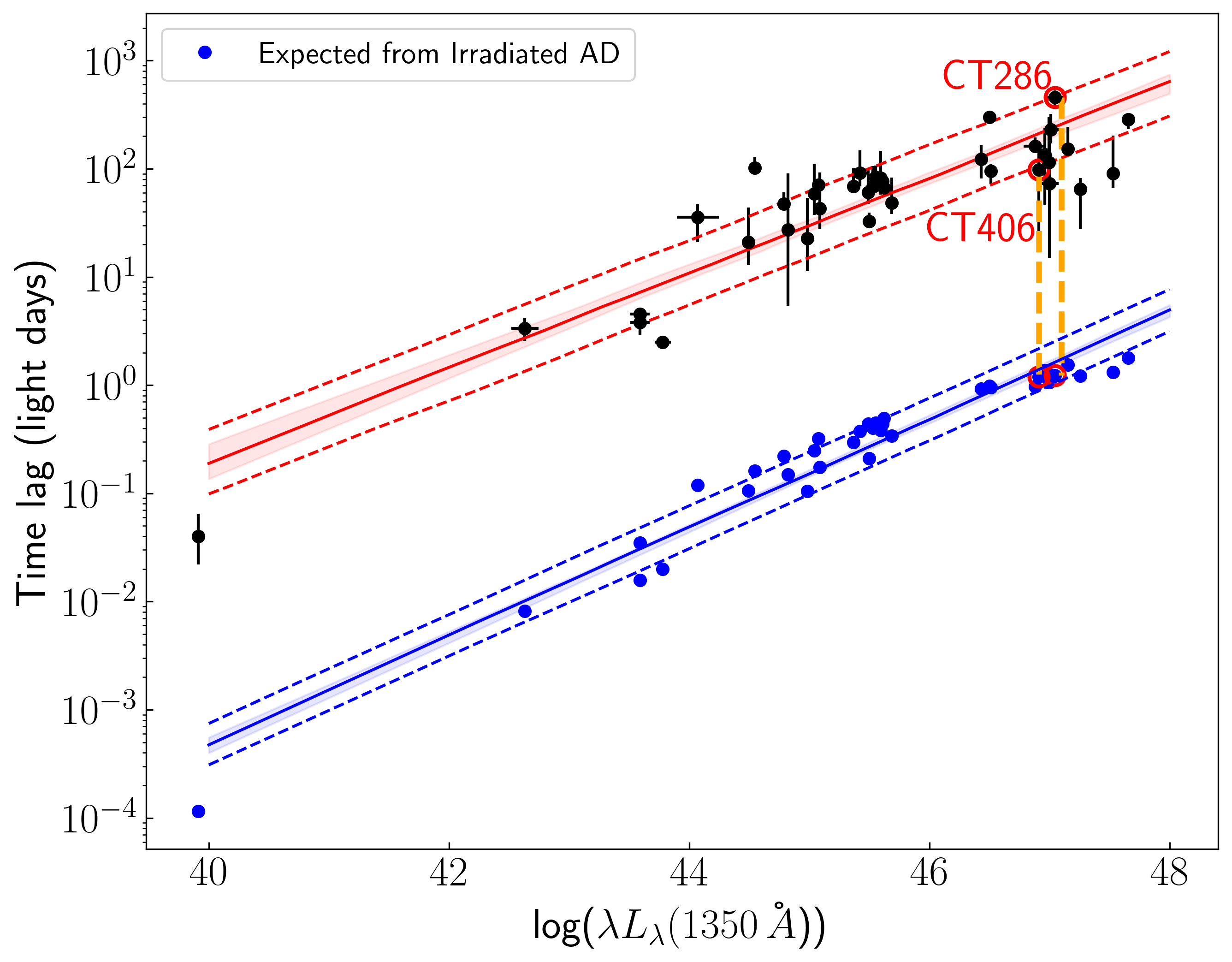}
    \caption{The $R_{\rm BLR}-L_{1350\text{\AA}}$ relation from \cite{2021ApJ...915..129K} (black) and the expected $R_{\rm AD}-L_{1350\text{\AA}}$ relation (blue). The solid red and blue lines indicate the mean values of the posterior probability distributions for both relationships. The shaded areas around these lines reflect the 1$\sigma$ uncertainty. The dashed lines delineate the mean predictions for the upper and lower bounds when the intrinsic scatter in the sources is considered. The positions of CT286 ($z = 2.556$) and CT406 ($z = 3.178$) are marked with red circles. The time lag is given in the rest frame.}
    \label{fig:adblrsize}
\end{figure}

As shown in \cite{2023MNRAS.522.2002P}, the measurement of quasar AD sizes can be significantly biased by external contaminants in the band-passes, e.g., BLR in the form of DCE and emission lines.
However, we note that the impact of these emissions can be reduced through a rigorous selection of filters.
This approach is used in Section \ref{sec4}, which makes the contamination by BLR emission lines negligible ($< 2\%$ for DCE and $\sim 2\%$ for He II and OIII lines).
Moreover, accounting for the extinction caused by the host galaxy and the AGN internal extinction is crucial to ensure accurate luminosity estimates.
Neglecting this factor can lead to a significant underestimation of luminosity, possibly by factors of 4 to 10 in the optical and UV spectra, respectively (\citealt{2017MNRAS.467..226G}).
We account for the contamination by the host galaxy by assuming, for simplicity, that the color is given by a Sa-galaxy profile (\citealt{1996ApJ...467...38K}) and contributes about 10-20\% of the total 5100\text{\AA} rest frame flux ($<2\%$ contribution at rest frame 1350\text{\AA}).
Finally, the nuclear extinction is added to the total flux, assuming the reddening curve of \cite{2007arXiv0711.1013G}. 
We emphasize that both the nuclear reddening and the contribution of the host galaxy represent constant components in the light curves that do not influence the estimation of the time delays. 
These corrections only affect the determination of the true AGN luminosity.

Here, we consider the time delay measured by the continuum near the \civ{} line at \(1647\,\text{\AA}\) to a reference time delay measured at \(1350\,\text{\AA}\).
We use \( \lambda L_{\lambda}(1350\text{\AA}) \) and \( M_{\rm BH} \) from \cite{2021ApJ...915..129K}, a sample that includes the sources whose \civ{} BLR sizes were recovered in \cite{2018ApJ...865...56L}, as described in Section \ref{sec2}.
We estimate \( \dot M \) assuming a bolometric luminosity correction \( L_{\rm Bol} = 10\lambda L_{\lambda}(5100\,\text{\AA}) \) (\citealt{2004MNRAS.352.1390M}) and a mass to radiation conversion efficiency \( \eta = L_{\rm Bol}/\dot M c^2 = 0.10 \) (\citealt{2009ApJ...690...20S}).
The expected AD time delays are therefore determined by the difference between the centroids of the transfer functions, \( \tau_{\rm cen} = \int \tau \Psi(\tau|\lambda) d\tau /\int \Psi(\tau|\lambda) d\tau, \) at \(1647\,\text{\AA}\) and \(1350\,\text{\AA}\) so that \( \tau_{\rm AD} = \tau_{\rm cen, 1647\,\text{\AA}} - \tau_{\rm cen, 1350\,\text{\AA}} \). 
In what follows, we refer to the \civ{} AD and BLR sizes as \( R_{\rm AD} = c\cdot\tau_{\rm AD} \) and \( R_{\rm BLR} = c\cdot\tau_{\rm BLR} \), respectively.

Figure \ref{fig:adblrsize} shows the predicted \( R_{\rm AD}-L_{1350\text{\AA}} \) along with the \( R_{\rm BLR}-L_{1350\text{\AA}} \) luminosity relation from \cite{2021ApJ...915..129K}.
We highlight the positions of quasars CT286 and CT406, which will be referenced in Section \ref{sec4} as benchmarks for evaluating the performance of the PRM observing campaign.

We performed a linear fit to both relationships, accounting for error measurements in luminosity, delays, and intrinsic scatter. 
The scatter was assumed to be normally distributed so that \( \log(R_{\rm BLR}/\text{lt-days}) \sim \mathcal{N}(\alpha \log(\lambda L_{\lambda}(1350\text{\AA})/10^{44} \, \text{erg\ s}^{-1}) + \kappa, \sigma_{\rm BLR}) \) and \( \log(R_{\rm AD}/\text{lt-days}) \sim \mathcal{N}(\beta \log(\lambda L_{\lambda}(1350\text{\AA})/10^{44} \, \text{erg\ s}^{-1}) + \gamma, \sigma_{\rm AD}) \).
We have no error measurements in the AD \civ{} delays, only in the luminosities. 
Therefore, we have assumed delay uncertainties of 10\%. 
This choice is justified in Section \ref{sec4}.
We derive the posterior probability distributions and the Bayesian evidence with the nested sampling Monte Carlo algorithm MLFriends (\citealt{2016S&C....26..383B,2019PASP..131j8005B}) using the UltraNest\footnote{\url{https://johannesbuchner.github.io/UltraNest/}} package (\citealt{2021JOSS....6.3001B}).
The best-fit results for the \( R_{\rm BLR}-L_{1350\text{\AA}} \) relation yield \( \alpha = 0.43^{+0.03}_{-0.04} \), \( \kappa = 1.04^{+0.08}_{-0.07} \), and \( \sigma_{\rm BLR} = 0.30^{+0.05}_{-0.04} \). For the \( R_{\rm AD}-L_{1350\text{\AA}} \) relation, the values are \( \beta = 0.50^{+0.02}_{-0.02} \), \( \gamma = -1.31^{+0.04}_{-0.04} \), and \( \sigma_{\rm AD} = 0.20^{+0.02}_{-0.02} \).
The fact that the $R_{\rm AD} - L$ relationship yields a slope of $\beta = 0.50$ is consistent with the expectations of the standard photoionization theory \citep{1999ApJ...526..579W, 2014AdSpR..54.1355N, 2021A&A...650A.154P}, which together with the smaller scatter $\sim 0.2$ dex compared to $\sim 0.3$ dex from the $R_{\rm BLR} - L$ relationship is an indication that the former is affected by contamination, e.g., by BLR scattering including DCE \citep{2022MNRAS.509.2637N, 2023MNRAS.522.2002P, 2023A&A...680A.102P, 2023A&A...670A.147J} and intrinsic reddening \citep{2007arXiv0711.1013G, 2023MNRAS.518..418H}. Furthermore, the \civ{} emission is dependent on the spectral shape of the ionizing continuum\footnote{the ionization potential for \civ{} is $\sim$64 eV, which corresponds to the soft X-ray region} of the accretion disk, which can develop additional anisotropy with increasing accretion rate \citep{2021A&A...650A.154P, 2023BoSAB..34..241P, 2023FrASS..1030103P}.

In the next section, we study the recovery of the delays under real observational conditions.

\section{Observing strategy}\label{sec4}

In this section, we outline the observation strategy and discuss the filters required to establish the $R_{\rm AD}-L_{1350\text{\AA}}$  relationship presented in Section \ref{sec3}. 
Our focus is primarily on the ESO La Silla 2.2-meter telescope. 
We highlight its capabilities as a monitoring instrument, emphasizing the comprehensive array of medium-band filters it offers. 
These features are particularly beneficial as they afford greater flexibility, allowing for coverage across a wider redshift range.
However, we note that the strategy can be extended to other meter-class telescopes, provided they have similar characteristics.

\subsection{La Silla 2.2-m MPG/ESO telescope}

Located at ESO La Silla observatory, the 2.2-meter telescope is equipped with the Wide Field Imager (WFI) camera, comprising a 4x2 mosaic of 2k $\times$ 4k CCDs yielding a field of view of $34\arcmin$ $\times$ $33\arcmin$ with a pixel scale of $0.238^{\prime\prime}$/pixel. 
 WFI has over 40 filters available, categorized into broad-band (FWHM $> 35$nm), medium-band (FWHM $> 15$nm), and narrow-band (FWHM $< 15$nm) covering between 340 nm up to  960 nm (\citealt{1999Msngr..95...15B}). 
Notably, the 26 existing medium-band filters provide a good balance, allowing efficient observations while minimizing BLR contamination.

Below, we present two examples of sources from Figure~\ref{fig:adblrsize}, the quasars CT286 and CT406. 
These two sources are included in the sample described in \cite{2018ApJ...865...56L}.
The selection is based on the availability of high-quality measurements of the \civ{} BLR sizes, luminosities, and black hole masses.

\begin{figure*}[!htb]
    \centering
    \includegraphics[width=\columnwidth]{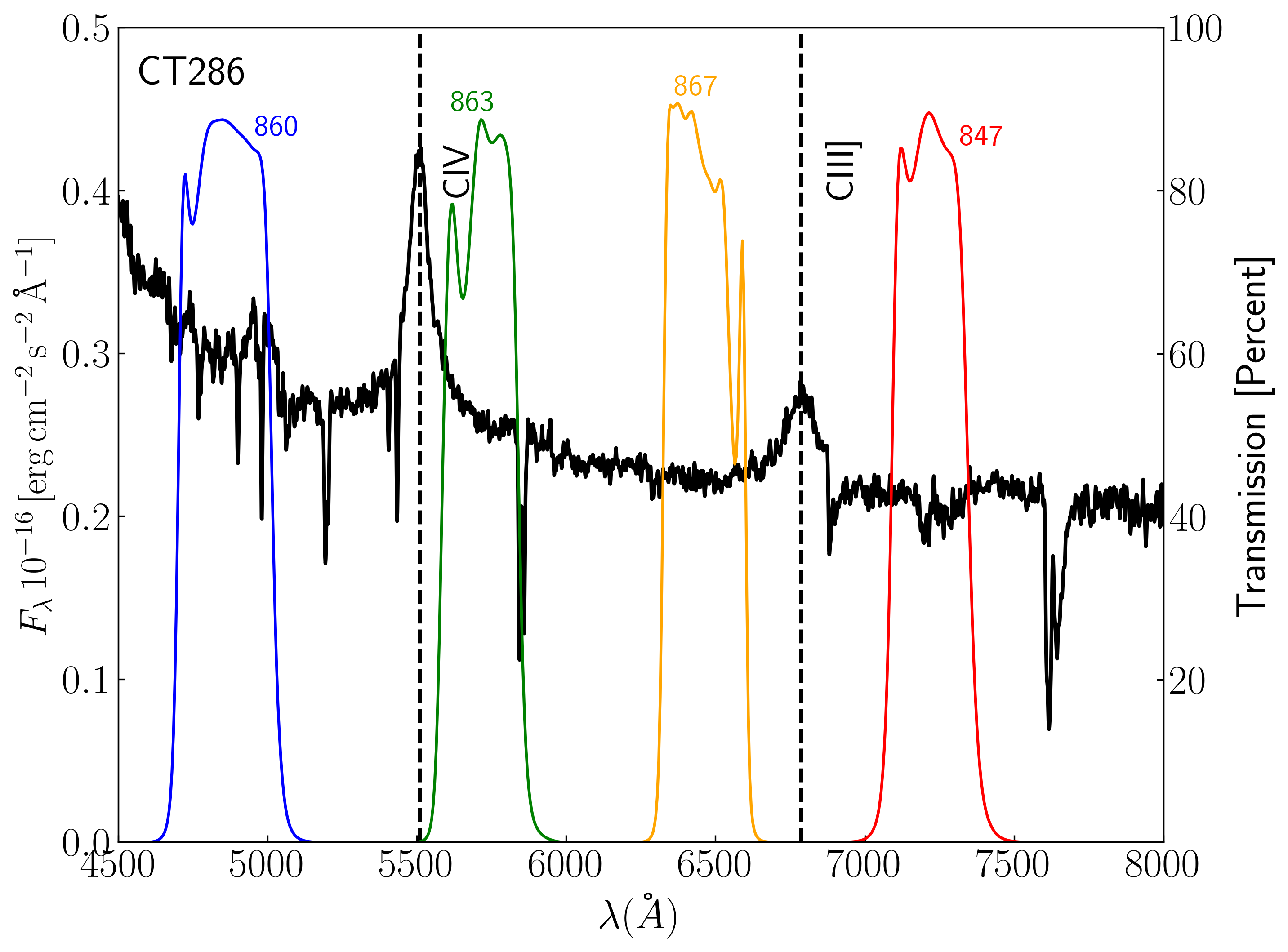}
    \includegraphics[width=\columnwidth]{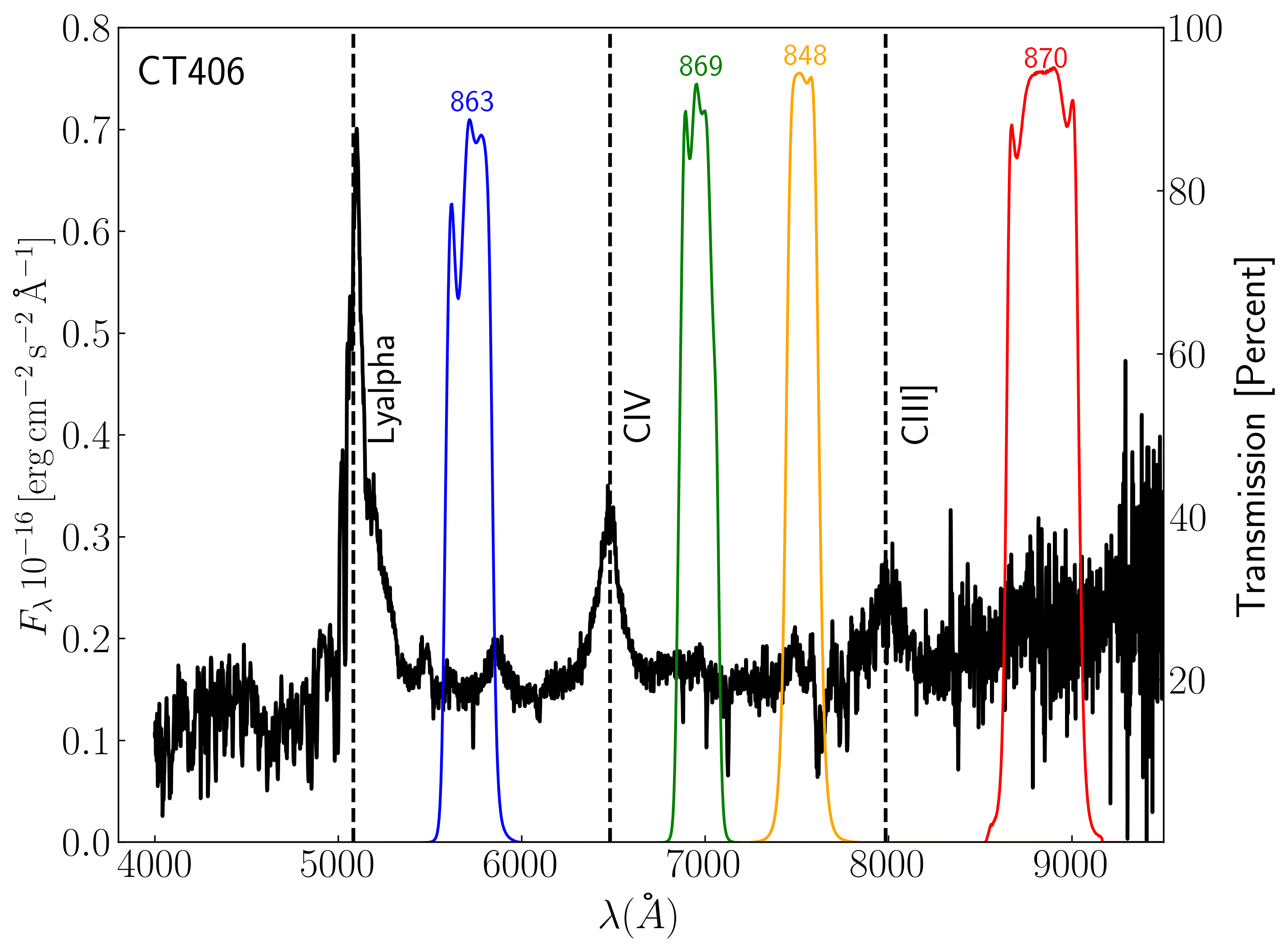}
    \caption{Spectrum of the sources CT286 (left) and CT406 (right) from \cite{2018ApJ...865...56L}. The colored line shows the transmission of the WFI medium-band filters, which were convolved with the quantum efficiency of the camera. The filters trace mainly the AGN emission-line free continuum variations around the \civ{} and \ciii{} emission lines.}
    \label{fig:selfil}
\end{figure*}

\subsection{CT286}

CT286 is located at a distance\footnote{Distances are derived from $z$ assuming a $\Lambda$-CDM cosmology with ${H_{0}=69.32\ \mathrm{km\ s^{-1}\ Mpc^{-1}}}$, $\Omega_{\Lambda}=0.7135$ and $\Omega_{m}=0.2865.$ \citep{2013ApJS..208...19H}.} of $\sim21$ Gpc with $z = 2.556$ (\citealt{1993RMxAA..25...51M}; \citealt{2018ApJ...865...56L}). \cite{2018ApJ...865...56L} reports \civ{}-based $M_{\rm BH} = (1.14\pm 0.23)\times 10^{9} M_{\odot}$ and $\lambda L_{\lambda}(1350\text{\AA}) = (1.12 \pm 0.18)\times 10^{47}{\mathrm{erg\ s^{-1}}}$. 
The spectra (provided by Paulina Lira, priv. comm.) and the selected WFI medium-band filters are shown in Figure \ref{fig:selfil}.
In this particular case, we have selected the WFI medium bands 860, 863, 867, and 847 with a central wavelength (and FWHM) at 4858 (315), 5714 (255), 6463 (277) and 7218 (257) $\text{\AA}$ respectively.
The WFI bands 847 and 867 trace continuum emission on the right and left sides of the \ciii{} line, and bands 863 and 860 trace the continuum on the left and right sides of \civ{}. 
The 860 band measures the rest frame flux at $1350\text{\AA}$ and is used as a reference band to measure the AD \civ{} delay, $\tau_{\rm AD}$, as described in Section \ref{sec3}.
The predicted time delay (observers frame) to the 860 band is 3.49, 6.74, and 10.15 days for bands 863, 867, and 847, respectively. Figure \ref{fig:tffs} shows the transfer functions and their centroids.

According to the WFI Exposure Time Calculator Version P113\footnote{\url{https://www.eso.org/observing/etc/bin/gen/form?INS.NAME=WFI+INS.MODE=imaging}}, at a brightness of $R = 16.89$, achieving an S/N $\sim 100$ requires on-source exposure times of 35, 45, 46 and 66 seconds for bands 860, 863, 867, and 847, respectively. The total on-source time for CT286 per night would be $\sim$4 minutes.

\subsection{CT406}

\begin{figure*}[!htb]
    \centering
    \includegraphics[width=\columnwidth]{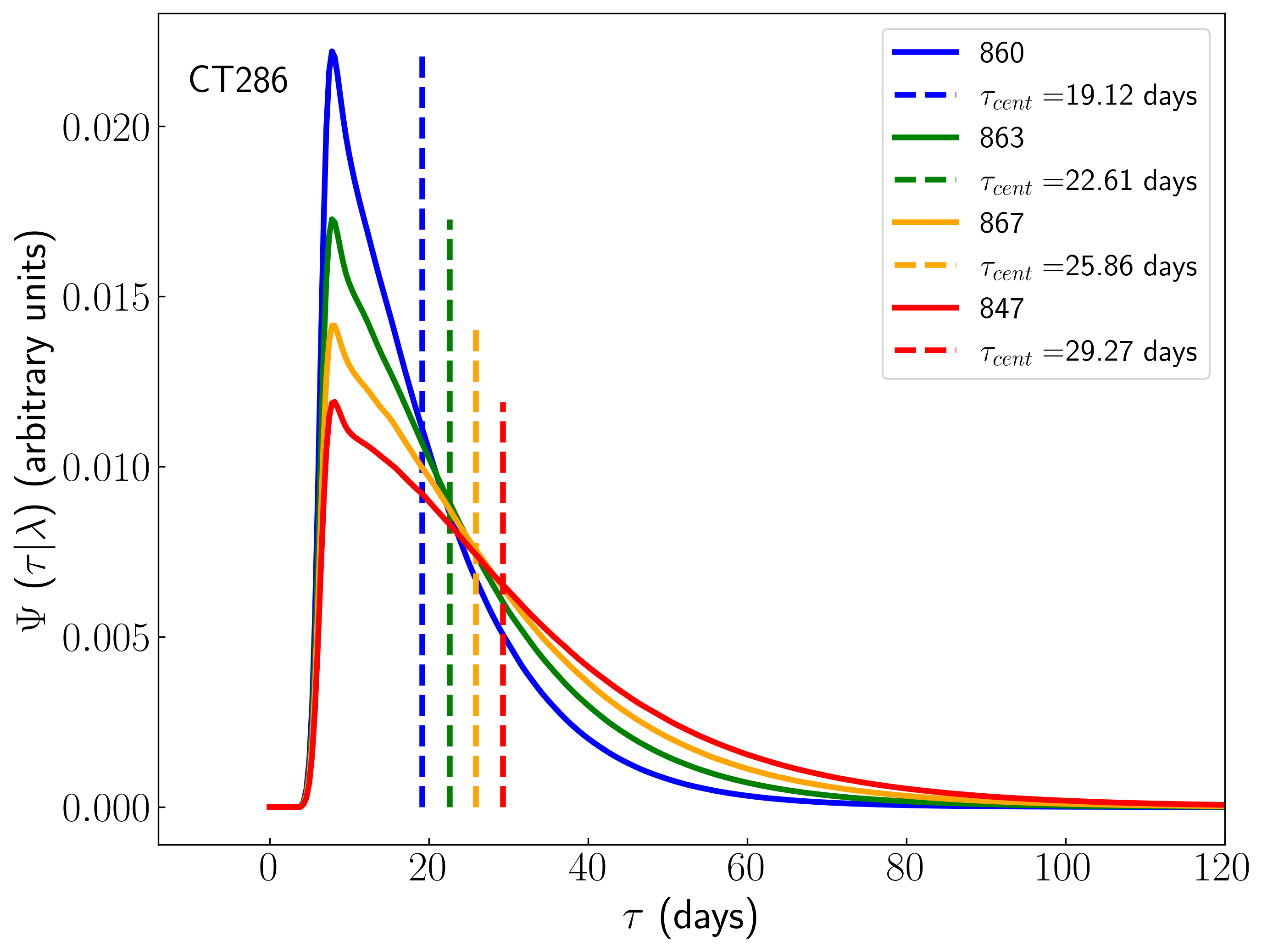}
    \includegraphics[width=\columnwidth]{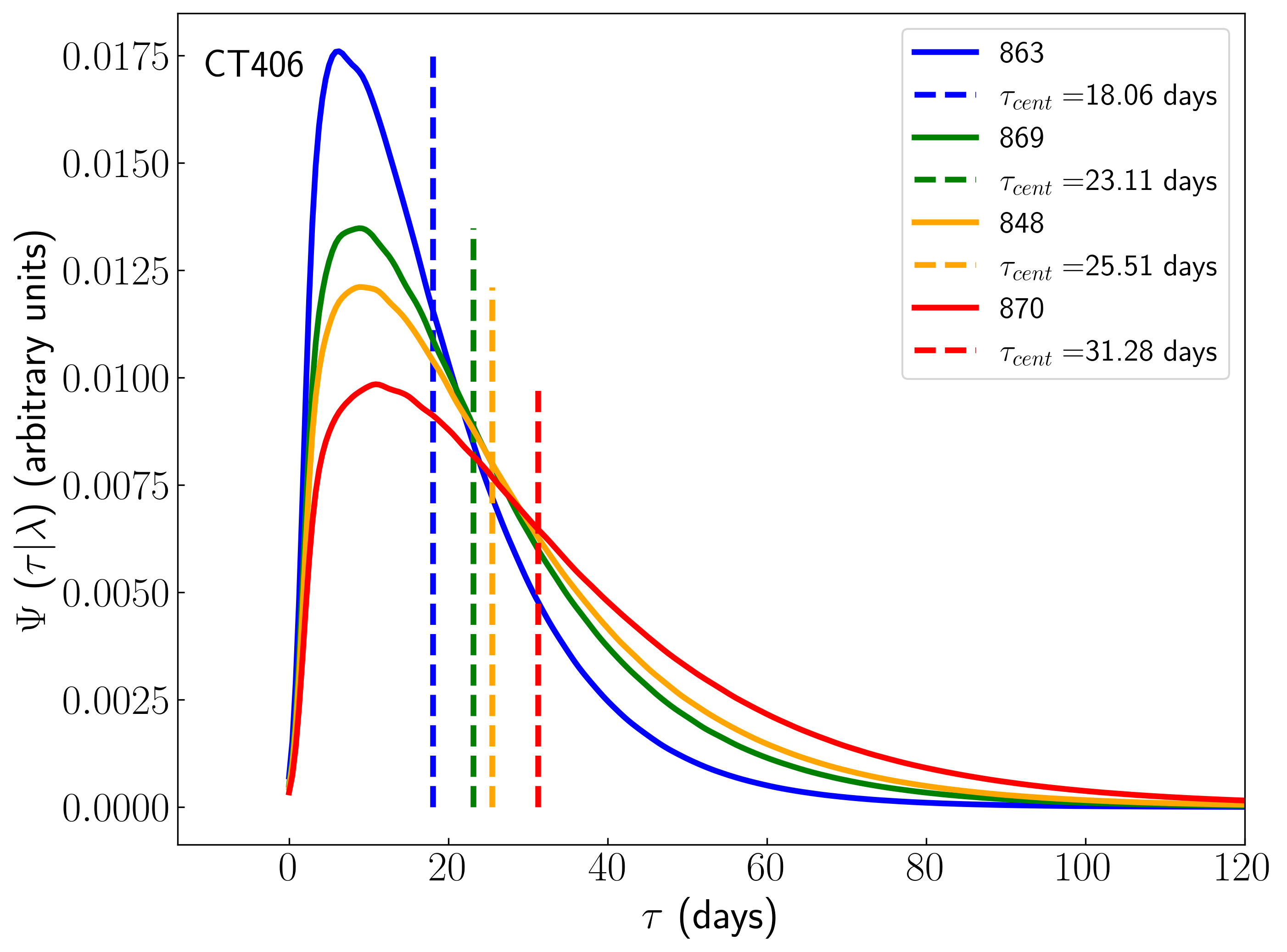}
    \caption{AD transfer functions for CT286 (left) and CT406 (right) sources. The transfer functions centroids are indicated in the observer frame and denoted by vertical dotted lines.}
    \label{fig:tffs}
\end{figure*}

CT406 is located at a distance of $\sim28$ Gpc with $z = 3.178$ (\citealt{1995RMxAA..31..119M}; \citealt{2018ApJ...865...56L}). \cite{2018ApJ...865...56L} reports \civ{}-based $M_{\rm BH} = (0.64\pm 0.42)\times 10^{9} M_{\odot}$ and $\lambda L_{\lambda}(1350\text{\AA}) = (8.13 \pm 0.75)\times 10^{46}{\mathrm{erg\ s^{-1}}}$. In this case, we have selected the WFI medium bands 863, 869, 848, and 870 (Figure \ref{fig:selfil}) with a central wavelength (and FWHM) at 5714 (255), 6963 (207), 7532 (183) and 8842 (397) $\text{\AA}$ respectively. The bands 870 and 848 trace continuum emission on the right and left sides of the \ciii{} line, and bands 869 and 863 trace the continuum on the left and right sides of \civ{}. 
The 863 band measures the rest frame flux at $1350\,\text{\AA}$, and it is used as the reference band to measure $\tau_{\rm AD}$. The predicted rest frame time delays given by the transfer functions (Figure \ref{fig:tffs}) are 5.05, 7.45, and 13.22 days for bands 869, 848, and 870, respectively.

At a brightness of $R = 17.66$, achieving an S/N $\sim 100$ requires on-source exposure times of 110, 161, 209, and 344 seconds for bands 863, 869, 848, and 870, respectively. 
The total on-source time for CT406 per night would be $\sim$14 minutes.

\begin{figure*}
\begin{tabular}{cc}
  \includegraphics[width=0.99\columnwidth]{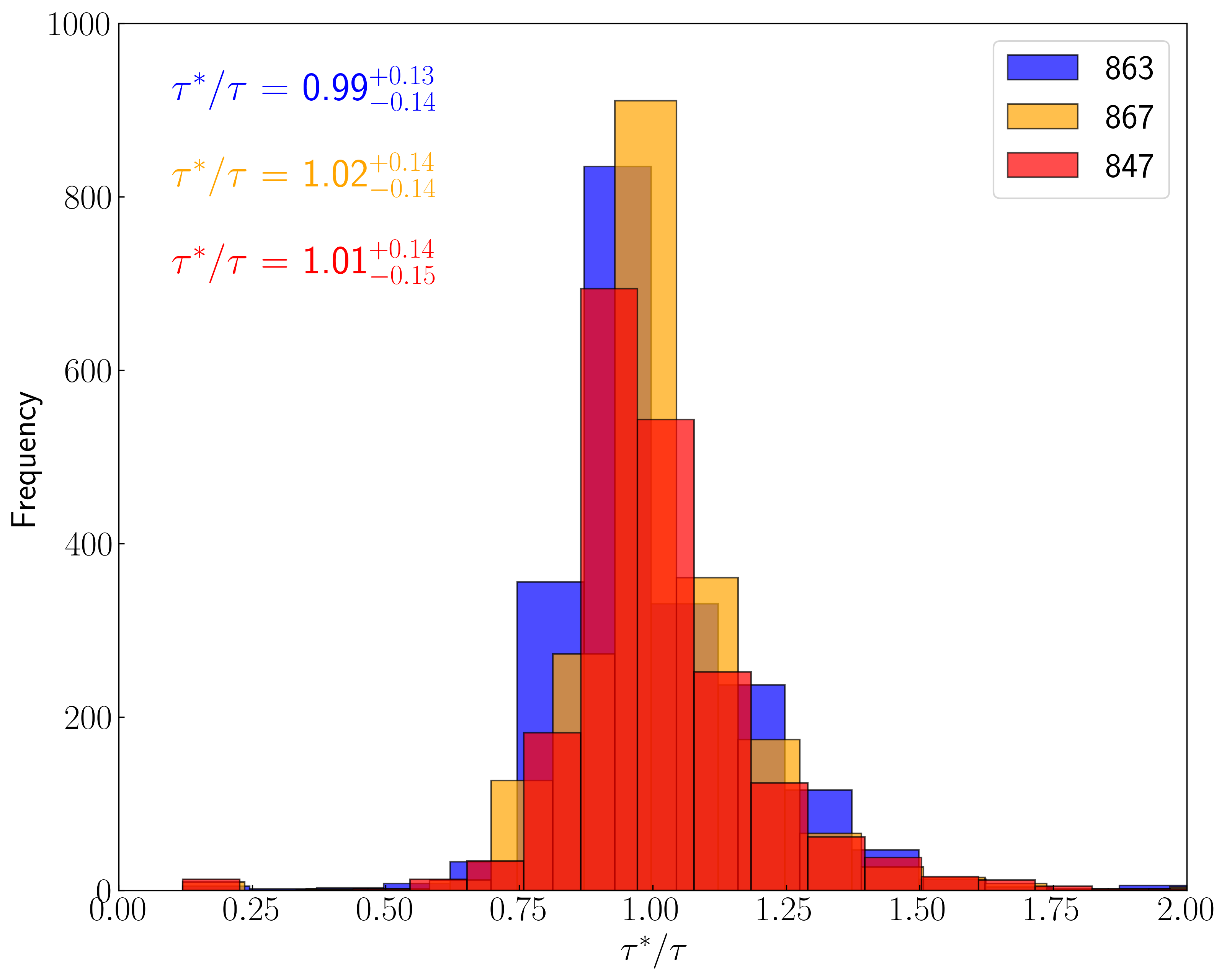} &   \includegraphics[width=0.99\columnwidth]{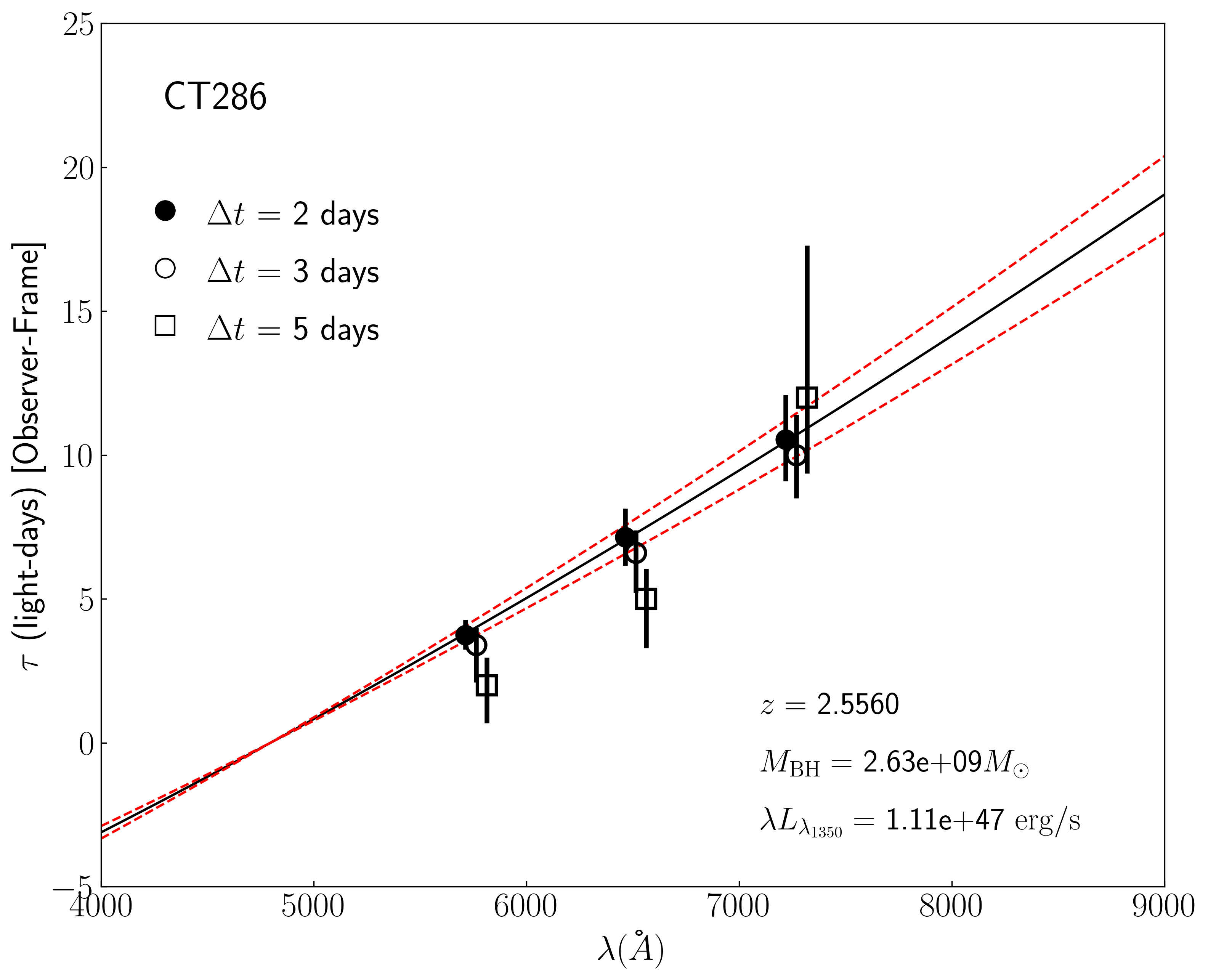}  \\ 
  \includegraphics[width=0.99\columnwidth]{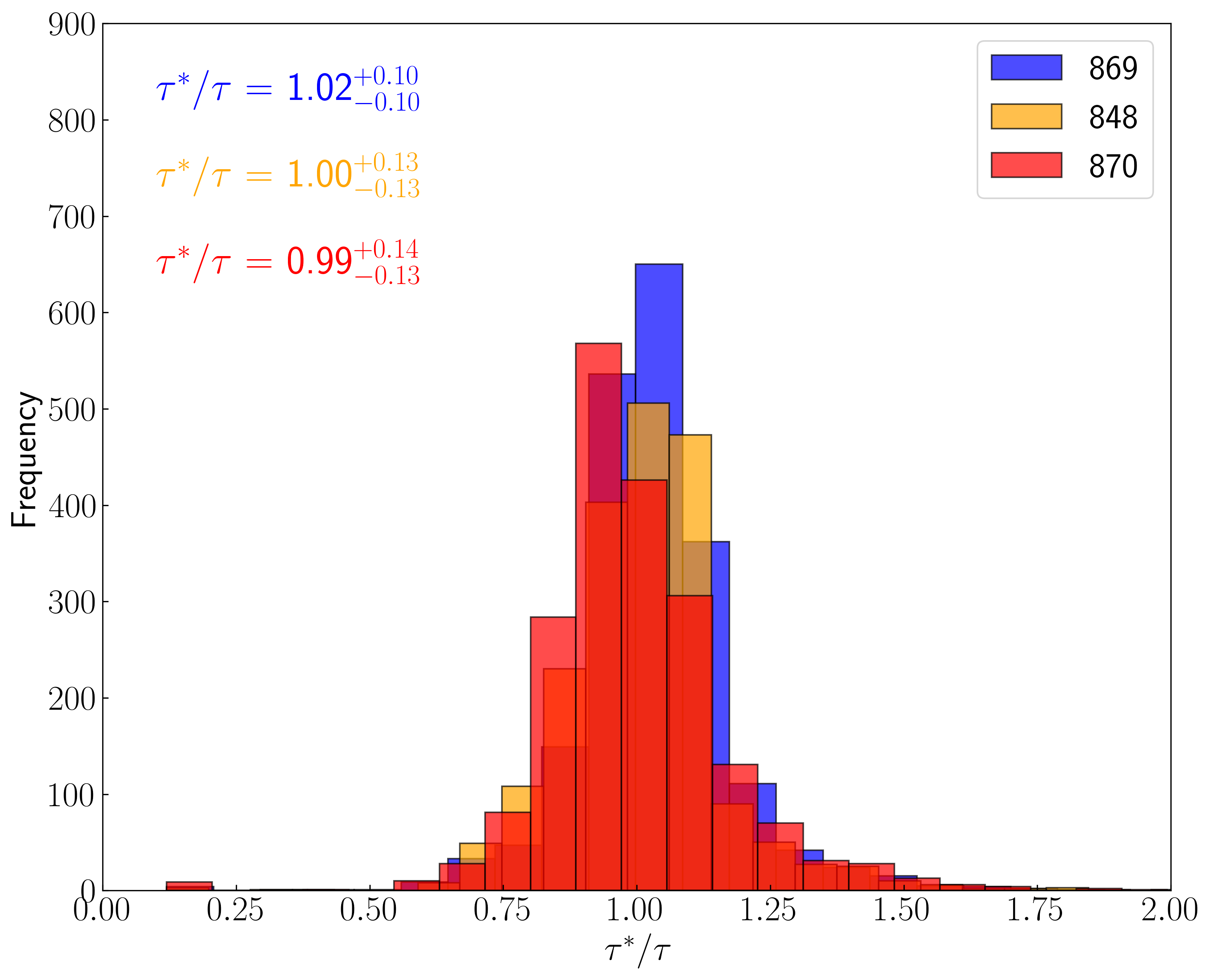} &   \includegraphics[width=0.99\columnwidth]{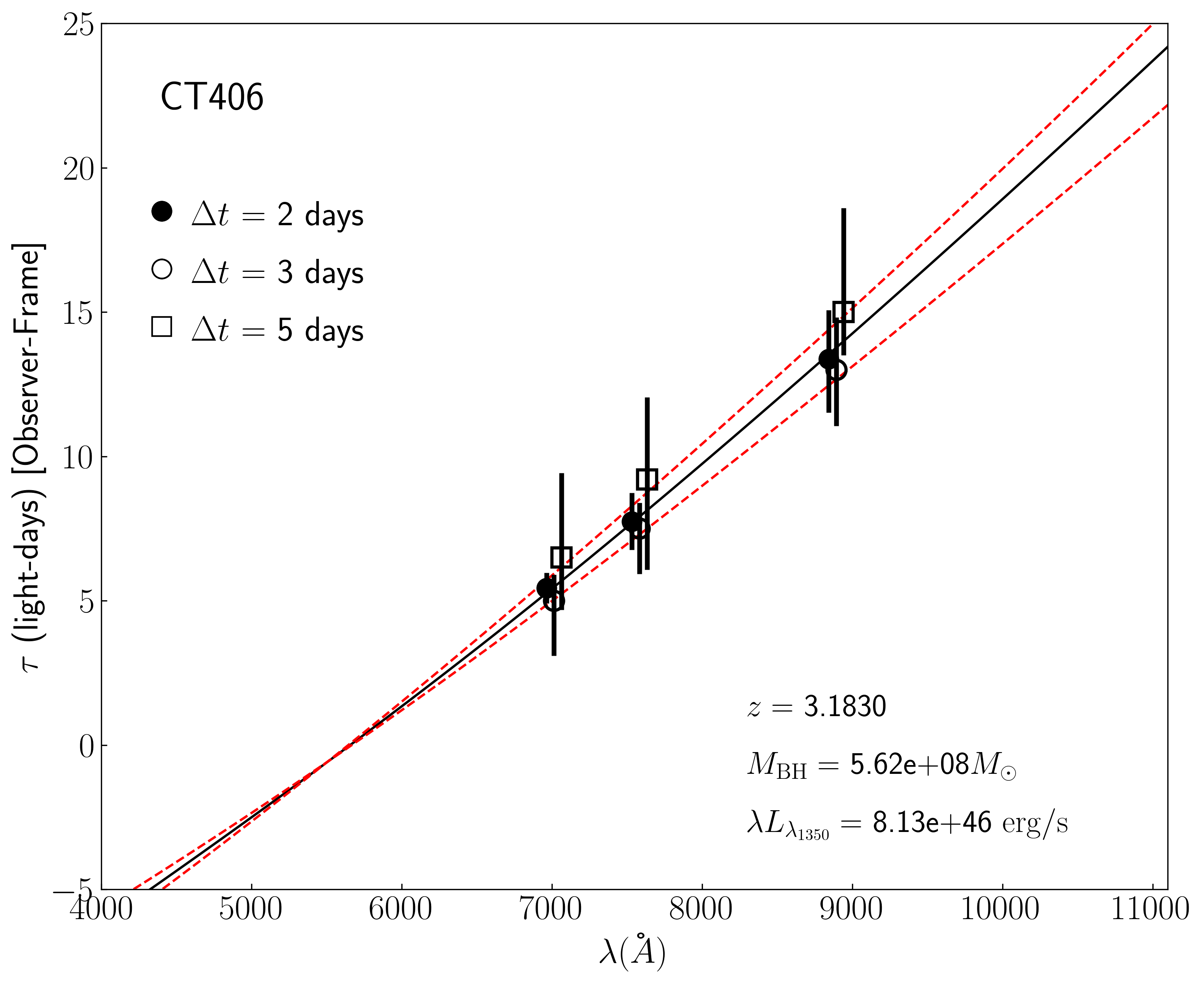}  \\
\end{tabular}
\caption{\textit{Left:} Recovered distributions of delays ($\tau^{*}$) for CT286 (top) and CT406 (bottom). \textit{Right:} time delay spectrum $\tau_{c}(\lambda)\propto \lambda^{4/3}$ (black line) as predicted from the transfer functions. The black circles represent the median values of the recovered distributions. The associated error bars reflect the upper and lower uncertainties, corresponding to the distributions' 16$^{\rm th}$ and 84$^{\rm th}$ percentiles, respectively. For comparison, we show the results for different sampling intervals using different symbols: filled circles for $\Delta t = 2$ days, empty circles for $\Delta t = 3$ days, and empty squares for $\Delta t = 5$ days. A small offset in $\lambda$ was used for better visualization. The dashed red lines show the delay spectrum obtained for a black hole mass with $30$\% uncertainty.}
\label{fig:distdelays}
\end{figure*}

\subsection{Time delays and the recovered CIV AD size - luminosity relation}

Using the simulations described in Section \ref{sec3}, we aim to quantify the accuracy with which the \civ{} AD time delays can be recovered. 
We use the quasars CT286 and CT406 as a reference for the physical parameter space used in the simulations.

The observed total fluxes are obtained from the convolution of the mixed AD+DCE+constant host galaxy and reddening components with the transmission curves of the WFI filters. 
The resulting light curves are resampled at average intervals, $\Delta t$, of 1, 2, 3, 4, and 5 days.
We added Gaussian noise and assumed measurement uncertainties at the level of $\sim 1\%$, which corresponds to a photometric signal-to-noise ratio of S/N $= 100$.
A total of 2000 noisy and resampled random light curves were generated for each source and the corresponding filters. 
This resulted in a total number of 16000 light curves that were used for the statistical analysis.
These light curves initially extend over 1000 days and are then subdivided into intervals of 180 days each. 
This subdivision reflects the typical duration of a 6-month observation campaign.

The time delays between pairs of light curves are calculated using the interpolated cross-correlation function (ICCF, \citealt{1987ApJS...65....1G}) with the latest probabilistic implementation (\citealt{2023A&A...674A..83P}).
The recovered distributions of delays ($\tau^{*}$) are shown in Figure \ref{fig:distdelays}. 
The distributions are shown for $\Delta t = 2.0$ days, where the delay is recovered with an accuracy of $\sim 15\%$ for CT286 and $\sim 10\%$ for CT406. 
The difference is primarily due to the higher redshift of CT406 ($z = 3.178$), which makes the recovery of the time delay more susceptible to time dilation effects (1+$z$), resulting in a larger deviation from Nyquist's theorem.
If the sampling is reduced to $\Delta t = 1.0$ day, the results are similar and show only a marginal performance improvement — an increase of 3\% for both sources (not shown). 

\begin{figure}
    \centering
    \includegraphics[width=\columnwidth]{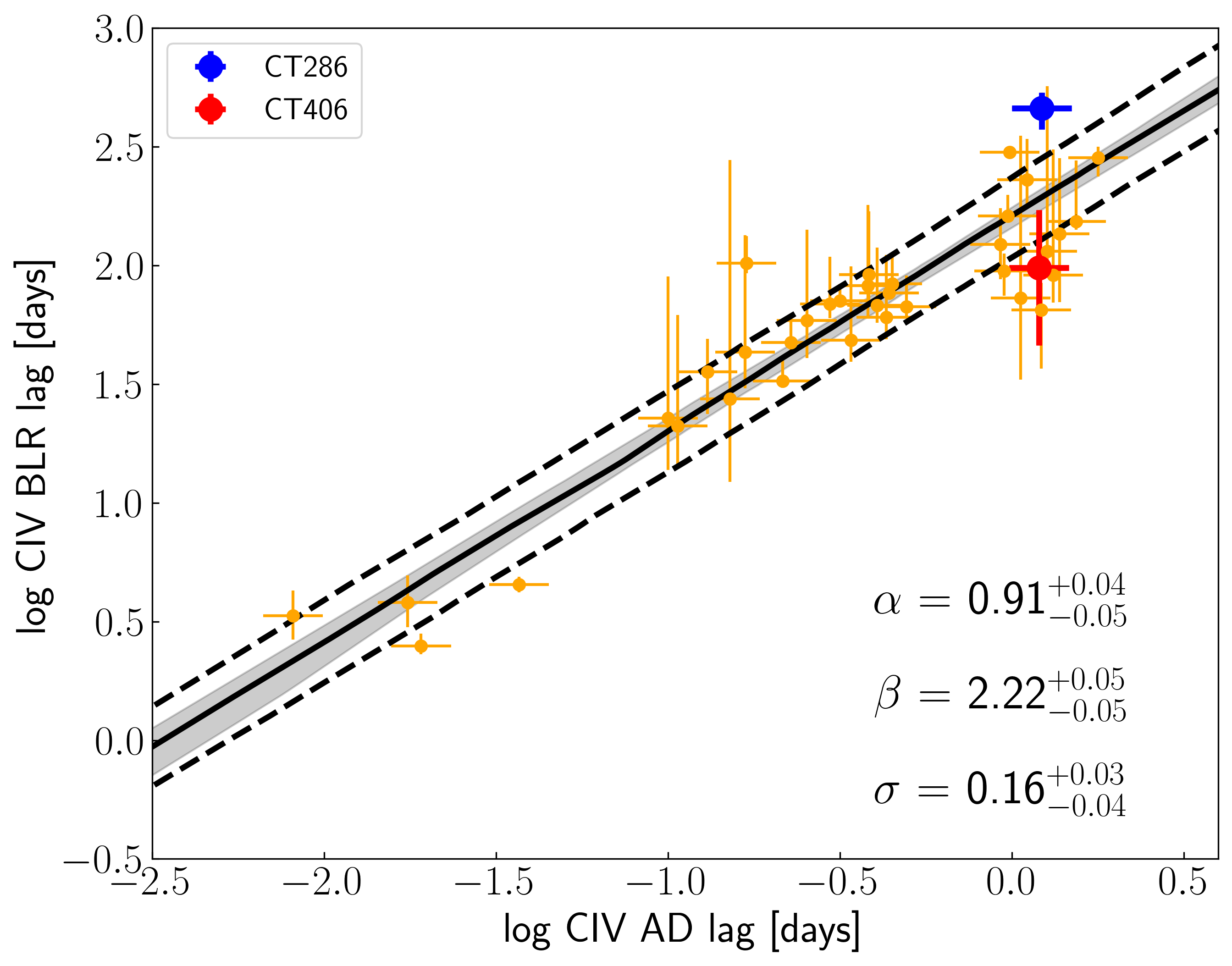}
    \caption{Recovered \( R_{\rm BLR} - R_{\rm AD} \) relationship. The solid black line indicates the mean value of the posterior probability distributions. The shaded area reflects the 1$\sigma$ uncertainty. The best-fit slope ($\alpha$), intercept ($\beta$), and scatter ($\sigma$) are reported with 1$\sigma$ uncertainties. The dotted lines delineate the mean predictions for the upper and lower bounds when the intrinsic scatter in the data is considered.
     Blue and red circles denote the positions of CT286 and CT406, respectively.}
    \label{fig:adblrsize_pred}
\end{figure}

The recovered time delay spectrum and the results obtained for $\Delta t = 2, 3$, and $5$ days are shown in Figure \ref{fig:distdelays} (right).
In the case of CT286, extending the time sampling interval to $\Delta t = 3$ days leads to a noticeable systematic bias  (up to about 10\%) toward small $\tau^{*}$ across all bands. 
In this setup, the accuracy in determining the true delay stands at 30\% for the shortest wavelength band (863 filter) and improves slightly for longer wavelengths, achieving 17\% for the 867 filter and 15\% for the 847 filter.
However, the situation degrades when the sampling interval is further increased to $\Delta t = 5$ days. 
In such a case, the bias towards smaller $\tau^{*}$ increases significantly, reaching up to 50\% for the shorter wavelength bands (863 and 867). 
Conversely, for the longer wavelength band (847), the bias shifts up to about 15\% toward larger values. 
Additionally, the overall precision in these bands markedly declines, dropping to 60\% for the 863 band and around 30\% for both the 867 and 847 bands.

CT406 exhibits similar behavior to CT286, as depicted in Figure \ref{fig:distdelays} (bottom right). 
However, as previously mentioned, better performance is observed for $\Delta t = 2$ days, where the delay is recovered with about 10\% precision across all bands. 
For $\Delta t = 3$ days, a systematic bias of about 7\% towards smaller values is noted. Under these conditions, the true delay is recovered with a precision of 28\%, 17\%, and 15\% for the 869, 848, and 870 bands, respectively. 
The accuracy deteriorates further for $\Delta t = 5$ days. In this scenario, the delay is recovered with about 35\% precision in all bands, accompanied by a clear systematic bias (up to about 20\%) towards larger $\tau^{*}$ in each band.

Overall, the results obtained for both objects show robust time-delay measurements, emphasizing that an appropriate choice of filters mitigates BLR contamination, which accounts for only up to about 2\% of the total flux in the passbands. Furthermore, it is unlikely that the systematic biases observed in all filters are due to external contamination, as the filters encompass different continuum regions of the spectrum. Instead, these biases are more plausibly due to temporal sampling and the convolution process itself.

The two sources, CT286 and CT406, belong to the brightest group of sources demonstrating the largest \civ{} BLR-AD time-lags as shown in Figure \ref{fig:adblrsize_pred}. Apart from these high-$z$, high-luminosity sources, the sources are primarily observed with the SDSS-RM project \citep{2019ApJ...887...38G}. In theory, one can also estimate the \civ{} AD time-delays for these sources. However, the SDSS-RM \civ{} sample is intrinsically faint, spanning an i-magnitude range between $\sim$19.5-21.5. With a 2.2m telescope, to reach an S/N of 100 a single on-target exposure with these medium band filters will take $\sim$4 hours. One would, therefore, need to try with a larger aperture instrument and capitalize on the high-$z$ sources that permit the recovery of the AD time delays without sacrificing the required cadence.

Based on the results of the time delay analysis, we can reasonably assume that with a sampling interval of $\Delta t = 2.0$ days, a delay accuracy of about 10\% to 15\% can be achieved over a 6-month monitoring campaign.

Figure \ref{fig:adblrsize_pred} shows the recovered  $R_{\rm AD} - R_{\rm BLR}$ relation, with the positions of CT286 and CT406 highlighted.

In a similar way to Section \ref{sec2}, we fitted the relation taking into account the intrinsic scatter and error measurements in \( R_{\rm BLR} \) and assuming a more conservative \( 20\% \) uncertainty in \( R_{\rm AD} \) with \( \log(R_{\rm BLR}/\text{lt-day}) \sim \mathcal{N}(\alpha \log(R_{\rm AD}/\text{lt-day}) + \beta, \sigma) \).
The results of the best fit are \( \alpha = 0.91^{+0.05}_{-0.04} \), \( \beta = 2.22^{+0.05}_{-0.05} \) and \( \sigma = 0.16^{+0.04}_{-0.03} \).
These results show that the size of the \civ{} BLR is on average about 150 times\footnote{{We caution the reader that this factor might change depending on the fraction of DCE contribution, as spectral decomposition might provide a lower limit to the actual DCE contribution to the bandpasses. The true DCE contribution varies among different objects, and more detailed photoionization simulations are necessary to accurately quantify its impact for individual sources. For instance, if one considers a 10\% contribution as given for NGC 5548 in the rest-frame 1300-2000 \AA\ range (\citealt{2019MNRAS.489.5284K}, Figure 9) to also apply to high-luminosity AGNs, the size of the \civ{} BLR would be on average about 40 times larger than that of the \civ{} AD.}} larger than that of the \civ{} AD, the latter measured at 1647\,\text{\AA}.
For example, for a quasar with an AD size of \( R_{\rm AD} = 1 \, \text{lt-day} \), we can predict a BLR size of \( R_{\rm BLR} = 165.9^{+36.2}_{-35.4} \, \text{lt-day} \), with an uncertainty of about \( 22\% \), considering the uncertainties of the parameters \( \alpha \), \( \beta \) and the intrinsic scatter \( \sigma \).

Taking into account the $\sim$5\% uncertainty in the FWHM measurements for the sources reported in \citet{2021ApJ...915..129K} (see their Table 6) and combining it with the 22\% uncertainty in the $R_{\rm BLR}$ scaling from our predictions, we calculate an overall uncertainty of $\sim$23\% in the $M_{\rm BH}$ estimates. 
This result is in agreement with the assumed 30\% uncertainty in $M_{\rm BH}$, as shown in Figure \ref{fig:distdelays}. 
However, this calculation does not take into account the uncertainties associated with the virial factor, which can contribute significantly to the error budget of the $M_{\rm BH}$ estimates \citep{2006A&A...456...75C, 2018NatAs...2...63M, 2019ApJ...882....4W, 2020ApJ...903..112D}. Assessing the \civ{}$\lambda$1549 profile, especially in high-accreting AGN, can be difficult and is a major contributor to the bias in the $M_{\rm BH}$ estimates. Previous works \citep[see e.g.,][]{2017A&A...608A.122S, 2020A&A...644A.175V, 2024Physi...6..216M} have found that the prominence of the outflowing component in \civ{} profile grows with the Eddington ratio of the source. More specifically, potential super-Eddington accretors exhibit a notable prevalence of outflows in their \civ{} emission line profiles yielding a shift of the emission profile by several thousand km/s to the rest frame \citep{2017A&A...608A.122S}. The shifted emission line signifies the presence of mildly ionized gas likely to escape the gravitational pull of the black hole and impart mechanical feedback to the host galaxy \citep{2016Ap&SS.361...29M}. It is crucial to disentangle the outflowing and virial components to estimate the virial black hole masses, otherwise leading to large uncertainties in the mass determination, and thus on Eddington ratios. We note that the choice of methodology can affect the mass measurements \citep[e.g.,][]{2018NatAs...2...63M, 2020ApJ...903..112D}, especially in the context of virial factors and choice of the FWHM or line dispersion ($\sigma_{\rm line}$).

\section{Summary and conclusions}

We used theoretical simulations to investigate the possibility of inferring the size of the \civ{} BLR from a measurement of the size of the AD. 
This is achieved through PRM, by measuring the time delay between the continuum variations near the \civ{} line at \(1647\,\text{\AA}\) and those at \(1350\,\text{\AA}\). The inferred BLR sizes can then be used to estimate black hole masses through the virial product.
We have provided the steps to carry out a successful PRM monitoring campaign of the AD on those high-redshift quasars where BLR \civ{} size measurements have already been performed. Our approach can be easily extended to other quasars at similar and higher redshifts where RM campaigns are yet to be made.
Here we have focused on the 2.2-metre telescope at ESO's La Silla Observatory, equipped with an extensive array of medium-band filters. 
These filters are remarkably efficient for PRM purposes. They are not too narrow to avoid excessive light loss and wide enough to effectively minimize contamination from BLR emissions.
Our results can be summarised as follows:
 
\begin{itemize}

      \item The size of the \civ{} BLR is, on average, about \textit{150} times larger, than that of the \civ{} AD as measured at 1647\,\text{\AA}. Therefore, a continuum size-luminosity relation can be recovered much more efficiently with PRM.

      \item A time sampling of 2 days, signal-to-noise ratio (S/N) of 100, and a BLR contribution of less than 2\% in the bandpasses can lead to a recovery of \civ{} continuum time delays with an accuracy of between 10\% and 15\% over a 6 months monitoring campaign. For a 2.2m telescope and bright quasars $16.8 < R < 17.8$ as the ones considered here, this requires about 5 to 15 minutes of total on-source exposure time per object.

      \item Assuming a model of an optically thick and geometrically thin accretion disk, the recovered time-delay spectrum agrees with the black hole masses derived with an accuracy of 30\% from the single-epoch BLR spectrum. A measurement of the size of the AD is used to estimate the BLR size with an accuracy of about 20\%. The overall uncertainty in the estimate of the black hole mass from the relation $R_{\rm AD}-L_{1350\text{\AA}}$ is about 23\%, not taking into account the uncertainties in the virial factor.

\end{itemize}

We have laid out a clear experiment to test whether one can infer BLR sizes from AD sizes.
It is important to carry out this experiment because it allows us to estimate BLR sizes 150x faster.
We note that this strategy can be applied to other meter-class telescopes with similarly efficient medium— or narrow-band filters.
Furthermore, our analysis has primarily focused on southern sources, as highlighted in the sample of \cite{2018ApJ...865...56L}, to take advantage of the unique observational opportunities offered by ESO's 2.2m telescope. 
However, the analysis can be extended to northern sources as well such as those described in \cite{2021ApJ...915..129K} and, given the efficiency of the method, to even higher redshift sources (e.g. $z \gtrsim$ 5). 
In this context, near-infrared instruments are crucial to enable the use of high-redshift AGN as probes to study BH evolution over cosmic time and constrain the cosmological parameters of our universe.

\begin{acknowledgements}
We thank Paulina Lira for sharing with us the spectra for CT286 and CT406. SP acknowledges the financial support of the Conselho Nacional de Desenvolvimento Científico e Tecnológico (CNPq) Fellowships 300936/2023-0 and 301628/2024-6. FPN gratefully acknowledges the generous and invaluable support of the Klaus Tschira Foundation. FPN acknowledges funding from the European Research Council (ERC) under the European Union's Horizon 2020 research and innovation program (grant agreement No 951549). This research has made use of the NASA/IPAC Extragalactic Database (NED) which is operated by the Jet Propulsion Laboratory, California Institute of Technology, under contract with the National Aeronautics and Space Administration. This research has made use of the SIMBAD database, operated at CDS, Strasbourg, France.
      
\end{acknowledgements}

%



\software{Matplotlib \citep{hunter2007}, 
        Numpy \citep{van2011}, MLFriends \citep{2016S&C....26..383B,2019PASP..131j8005B}, Scipy \citep{virtanen2020}, UltraNest \citep{2021JOSS....6.3001B}, GPCC \citep{2023A&A...674A..83P}}





\bibliography{references}{}

\begin{thebibliography}{}
\expandafter\ifx\csname natexlab\endcsname\relax\def\natexlab#1{#1}\fi
\providecommand{\url}[1]{\href{#1}{#1}}
\providecommand{\dodoi}[1]{doi:~\href{http://doi.org/#1}{\nolinkurl{#1}}}
\providecommand{\doeprint}[1]{\href{http://ascl.net/#1}{\nolinkurl{http://ascl.net/#1}}}
\providecommand{\doarXiv}[1]{\href{https://arxiv.org/abs/#1}{\nolinkurl{https://arxiv.org/abs/#1}}}

\bibitem[{{Baade} {et~al.}(1999){Baade}, {Meisenheimer}, {Iwert}, {Alonso},
  {Augusteijn}, {Beletic}, {Bellemann}, {Benesch}, {B{\"o}hm}, {B{\"o}hnhardt},
  {Brewer}, {Deiries}, {Delabre}, {Donaldson}, {Dupuy}, {Franke}, {Gerdes},
  {Gilliotte}, {Grimm}, {Haddad}, {Hess}, {Ihle}, {Klein}, {Lenzen}, {Lizon},
  {Mancini}, {M{\"u}nch}, {Pizarro}, {Prado}, {Rahmer}, {Reyes}, {Richardson},
  {Robledo}, {Sanchez}, {Silber}, {Sinclaire}, {Wackermann}, \&
  {Zaggia}}]{1999Msngr..95...15B}
{Baade}, D., {Meisenheimer}, K., {Iwert}, O., {et~al.} 1999, The Messenger, 95,
  15

\bibitem[{{Bentz} {et~al.}(2009){Bentz}, {Walsh}, {Barth}, {Baliber},
  {Bennert}, {Canalizo}, {Filippenko}, {Ganeshalingam}, {Gates}, {Greene},
  {Hidas}, {Hiner}, {Lee}, {Li}, {Malkan}, {Minezaki}, {Sakata}, {Serduke},
  {Silverman}, {Steele}, {Stern}, {Street}, {Thornton}, {Treu}, {Wang}, {Woo},
  \& {Yoshii}}]{2009ApJ...705..199B}
{Bentz}, M.~C., {Walsh}, J.~L., {Barth}, A.~J., {et~al.} 2009, \apj, 705, 199,
  \dodoi{10.1088/0004-637X/705/1/199}

\bibitem[{{Bentz} {et~al.}(2013){Bentz}, {Denney}, {Grier}, {Barth},
  {Peterson}, {Vestergaard}, {Bennert}, {Canalizo}, {De Rosa}, {Filippenko},
  {Gates}, {Greene}, {Li}, {Malkan}, {Pogge}, {Stern}, {Treu}, \&
  {Woo}}]{2013ApJ...767..149B}
{Bentz}, M.~C., {Denney}, K.~D., {Grier}, C.~J., {et~al.} 2013, \apj, 767, 149,
  \dodoi{10.1088/0004-637X/767/2/149}

\bibitem[{{Buchner}(2016)}]{2016S&C....26..383B}
{Buchner}, J. 2016, Statistics and Computing, 26, 383,
  \dodoi{10.1007/s11222-014-9512-y}

\bibitem[{{Buchner}(2019)}]{2019PASP..131j8005B}
---. 2019, \pasp, 131, 108005, \dodoi{10.1088/1538-3873/aae7fc}

\bibitem[{{Buchner}(2021)}]{2021JOSS....6.3001B}
---. 2021, The Journal of Open Source Software, 6, 3001,
  \dodoi{10.21105/joss.03001}

\bibitem[{{Cackett} {et~al.}(2021){Cackett}, {Bentz}, \&
  {Kara}}]{2021iSci...24j2557C}
{Cackett}, E.~M., {Bentz}, M.~C., \& {Kara}, E. 2021, iScience, 24, 102557,
  \dodoi{10.1016/j.isci.2021.102557}

\bibitem[{{Cackett} {et~al.}(2018){Cackett}, {Chiang}, {McHardy}, {Edelson},
  {Goad}, {Horne}, \& {Korista}}]{2018ApJ...857...53C}
{Cackett}, E.~M., {Chiang}, C.-Y., {McHardy}, I., {et~al.} 2018, \apj, 857, 53,
  \dodoi{10.3847/1538-4357/aab4f7}

\bibitem[{{Cackett} {et~al.}(2007){Cackett}, {Horne}, \&
  {Winkler}}]{2007MNRAS.380..669C}
{Cackett}, E.~M., {Horne}, K., \& {Winkler}, H. 2007, \mnras, 380, 669,
  \dodoi{10.1111/j.1365-2966.2007.12098.x}

\bibitem[{{Cao} {et~al.}(2023){Cao}, {Zaja{\v{c}}ek}, {Czerny}, {Panda}, \&
  {Ratra}}]{2023arXiv230916516C}
{Cao}, S., {Zaja{\v{c}}ek}, M., {Czerny}, B., {Panda}, S., \& {Ratra}, B. 2023,
  arXiv e-prints, arXiv:2309.16516, \dodoi{10.48550/arXiv.2309.16516}

\bibitem[{{Caplar} {et~al.}(2017){Caplar}, {Lilly}, \&
  {Trakhtenbrot}}]{2017ApJ...834..111C}
{Caplar}, N., {Lilly}, S.~J., \& {Trakhtenbrot}, B. 2017, \apj, 834, 111,
  \dodoi{10.3847/1538-4357/834/2/111}

\bibitem[{{Chartas} {et~al.}(2016){Chartas}, {Rhea}, {Kochanek}, {Dai},
  {Morgan}, {Blackburne}, {Chen}, {Mosquera}, \&
  {MacLeod}}]{2016AN....337..356C}
{Chartas}, G., {Rhea}, C., {Kochanek}, C., {et~al.} 2016, Astronomische
  Nachrichten, 337, 356, \dodoi{10.1002/asna.201612313}

\bibitem[{{Chelouche} {et~al.}(2019){Chelouche}, {Pozo Nu{\~n}ez}, \&
  {Kaspi}}]{2019NatAs...3..251C}
{Chelouche}, D., {Pozo Nu{\~n}ez}, F., \& {Kaspi}, S. 2019, Nature Astronomy,
  3, 251, \dodoi{10.1038/s41550-018-0659-x}

\bibitem[{{Collier} \& {Peterson}(2001)}]{2001ApJ...555..775C}
{Collier}, S., \& {Peterson}, B.~M. 2001, \apj, 555, 775,
  \dodoi{10.1086/321517}

\bibitem[{{Collier} {et~al.}(1998){Collier}, {Horne}, {Kaspi}, {Netzer},
  {Peterson}, {Wanders}, {Alexander}, {Bertram}, {Comastri}, {Gaskell},
  {Malkov}, {Maoz}, {Mignoli}, {Pogge}, {Pronik}, {Sergeev}, {Snedden},
  {Stirpe}, {Bochkarev}, {Burenkov}, {Shapovalova}, \&
  {Wagner}}]{1998ApJ...500..162C}
{Collier}, S.~J., {Horne}, K., {Kaspi}, S., {et~al.} 1998, \apj, 500, 162,
  \dodoi{10.1086/305720}

\bibitem[{{Collin} {et~al.}(2006){Collin}, {Kawaguchi}, {Peterson}, \&
  {Vestergaard}}]{2006A&A...456...75C}
{Collin}, S., {Kawaguchi}, T., {Peterson}, B.~M., \& {Vestergaard}, M. 2006,
  \aap, 456, 75, \dodoi{10.1051/0004-6361:20064878}

\bibitem[{{Czerny} \& {Elvis}(1987)}]{1987ApJ...321..305C}
{Czerny}, B., \& {Elvis}, M. 1987, \apj, 321, 305, \dodoi{10.1086/165630}

\bibitem[{{Czerny} \& {Hryniewicz}(2011)}]{2011A&A...525L...8C}
{Czerny}, B., \& {Hryniewicz}, K. 2011, \aap, 525, L8,
  \dodoi{10.1051/0004-6361/201016025}

\bibitem[{{Dalla Bont{\`a}} {et~al.}(2020){Dalla Bont{\`a}}, {Peterson},
  {Bentz}, {Brandt}, {Ciroi}, {De Rosa}, {Fonseca Alvarez}, {Grier}, {Hall},
  {Hern{\'a}ndez Santisteban}, {Ho}, {Homayouni}, {Horne}, {Kochanek}, {Li},
  {Morelli}, {Pizzella}, {Pogge}, {Schneider}, {Shen}, {Trump}, \&
  {Vestergaard}}]{2020ApJ...903..112D}
{Dalla Bont{\`a}}, E., {Peterson}, B.~M., {Bentz}, M.~C., {et~al.} 2020, \apj,
  903, 112, \dodoi{10.3847/1538-4357/abbc1c}

\bibitem[{{Davidson} \& {Netzer}(1979)}]{1979RvMP...51..715D}
{Davidson}, K., \& {Netzer}, H. 1979, Reviews of Modern Physics, 51, 715,
  \dodoi{10.1103/RevModPhys.51.715}

\bibitem[{{Du} \& {Wang}(2019)}]{2019ApJ...886...42D}
{Du}, P., \& {Wang}, J.-M. 2019, \apj, 886, 42,
  \dodoi{10.3847/1538-4357/ab4908}

\bibitem[{{Edelson} {et~al.}(2015){Edelson}, {Gelbord}, {Horne}, {McHardy},
  {Peterson}, {Ar{\'e}valo}, {Breeveld}, {De Rosa}, {Evans}, {Goad}, {Kriss},
  {Brandt}, {Gehrels}, {Grupe}, {Kennea}, {Kochanek}, {Nousek}, {Papadakis},
  {Siegel}, {Starkey}, {Uttley}, {Vaughan}, {Young}, {Barth}, {Bentz},
  {Brewer}, {Crenshaw}, {Dalla Bont{\`a}}, {De Lorenzo-C{\'a}ceres}, {Denney},
  {Dietrich}, {Ely}, {Fausnaugh}, {Grier}, {Hall}, {Kaastra}, {Kelly},
  {Korista}, {Lira}, {Mathur}, {Netzer}, {Pancoast}, {Pei}, {Pogge},
  {Schimoia}, {Treu}, {Vestergaard}, {Villforth}, {Yan}, \&
  {Zu}}]{2015ApJ...806..129E}
{Edelson}, R., {Gelbord}, J.~M., {Horne}, K., {et~al.} 2015, \apj, 806, 129,
  \dodoi{10.1088/0004-637X/806/1/129}

\bibitem[{{Fausnaugh} {et~al.}(2016){Fausnaugh}, {Denney}, {Barth}, {Bentz},
  {Bottorff}, {Carini}, {Croxall}, {De Rosa}, {Goad}, {Horne}, {Joner},
  {Kaspi}, {Kim}, {Klimanov}, {Kochanek}, {Leonard}, {Netzer}, {Peterson},
  {Schn{\"u}lle}, {Sergeev}, {Vestergaard}, {Zheng}, {Zu}, {Anderson},
  {Ar{\'e}valo}, {Bazhaw}, {Borman}, {Boroson}, {Brandt}, {Breeveld}, {Brewer},
  {Cackett}, {Crenshaw}, {Dalla Bont{\`a}}, {De Lorenzo-C{\'a}ceres},
  {Dietrich}, {Edelson}, {Efimova}, {Ely}, {Evans}, {Filippenko}, {Flatland},
  {Gehrels}, {Geier}, {Gelbord}, {Gonzalez}, {Gorjian}, {Grier}, {Grupe},
  {Hall}, {Hicks}, {Horenstein}, {Hutchison}, {Im}, {Jensen}, {Jones},
  {Kaastra}, {Kelly}, {Kennea}, {Kim}, {Korista}, {Kriss}, {Lee}, {Lira},
  {MacInnis}, {Manne-Nicholas}, {Mathur}, {McHardy}, {Montouri}, {Musso},
  {Nazarov}, {Norris}, {Nousek}, {Okhmat}, {Pancoast}, {Papadakis}, {Parks},
  {Pei}, {Pogge}, {Pott}, {Rafter}, {Rix}, {Saylor}, {Schimoia}, {Siegel},
  {Spencer}, {Starkey}, {Sung}, {Teems}, {Treu}, {Turner}, {Uttley},
  {Villforth}, {Weiss}, {Woo}, {Yan}, \& {Young}}]{2016ApJ...821...56F}
{Fausnaugh}, M.~M., {Denney}, K.~D., {Barth}, A.~J., {et~al.} 2016, \apj, 821,
  56, \dodoi{10.3847/0004-637X/821/1/56}

\bibitem[{{Frank} {et~al.}(2002){Frank}, {King}, \&
  {Raine}}]{2002apa..book.....F}
{Frank}, J., {King}, A., \& {Raine}, D.~J. 2002, {Accretion Power in
  Astrophysics: Third Edition}

\bibitem[{{Gaskell}(2017)}]{2017MNRAS.467..226G}
{Gaskell}, C.~M. 2017, \mnras, 467, 226, \dodoi{10.1093/mnras/stx094}

\bibitem[{{Gaskell} {et~al.}(2023){Gaskell}, {Anderson}, {Birmingham}, \&
  {Ghosh}}]{2023MNRAS.519.4082G}
{Gaskell}, C.~M., {Anderson}, F.~C., {Birmingham}, S.~{\'A}., \& {Ghosh}, S.
  2023, \mnras, 519, 4082, \dodoi{10.1093/mnras/stac3333}

\bibitem[{{Gaskell} \& {Benker}(2007)}]{2007arXiv0711.1013G}
{Gaskell}, C.~M., \& {Benker}, A.~J. 2007, arXiv e-prints, arXiv:0711.1013,
  \dodoi{10.48550/arXiv.0711.1013}

\bibitem[{{Gaskell} \& {Peterson}(1987)}]{1987ApJS...65....1G}
{Gaskell}, C.~M., \& {Peterson}, B.~M. 1987, \apjs, 65, 1,
  \dodoi{10.1086/191216}

\bibitem[{{Gianniotis} {et~al.}(2022){Gianniotis}, {Pozo Nu{\~n}ez}, \&
  {Polsterer}}]{2022A&A...657A.126G}
{Gianniotis}, N., {Pozo Nu{\~n}ez}, F., \& {Polsterer}, K.~L. 2022, \aap, 657,
  A126, \dodoi{10.1051/0004-6361/202141710}

\bibitem[{{Giveon} {et~al.}(1999){Giveon}, {Maoz}, {Kaspi}, {Netzer}, \&
  {Smith}}]{1999MNRAS.306..637G}
{Giveon}, U., {Maoz}, D., {Kaspi}, S., {Netzer}, H., \& {Smith}, P.~S. 1999,
  \mnras, 306, 637, \dodoi{10.1046/j.1365-8711.1999.02556.x}

\bibitem[{{Gonz{\'a}lez-Buitrago} {et~al.}(2023){Gonz{\'a}lez-Buitrago},
  {Garc{\'\i}a-D{\'\i}az}, {Pozo Nu{\~n}ez}, \& {Guo}}]{2023MNRAS.525.4524G}
{Gonz{\'a}lez-Buitrago}, D.~H., {Garc{\'\i}a-D{\'\i}az}, M.~T., {Pozo
  Nu{\~n}ez}, F., \& {Guo}, H. 2023, \mnras, 525, 4524,
  \dodoi{10.1093/mnras/stad2483}

\bibitem[{{Grier} {et~al.}(2019){Grier}, {Shen}, {Horne}, {Brandt}, {Trump},
  {Hall}, {Kinemuchi}, {Starkey}, {Schneider}, {Ho}, {Homayouni}, {I-Hsiu Li},
  {McGreer}, {Peterson}, {Bizyaev}, {Chen}, {Dawson}, {Eftekharzadeh}, {Guo},
  {Jia}, {Jiang}, {Kneib}, {Li}, {Li}, {Nie}, {Oravetz}, {Oravetz}, {Pan},
  {Petitjean}, {Ponder}, {Rogerson}, {Vivek}, {Zhang}, \&
  {Zou}}]{2019ApJ...887...38G}
{Grier}, C.~J., {Shen}, Y., {Horne}, K., {et~al.} 2019, \apj, 887, 38,
  \dodoi{10.3847/1538-4357/ab4ea5}

\bibitem[{{Hawkins}(2007)}]{2007A&A...462..581H}
{Hawkins}, M.~R.~S. 2007, \aap, 462, 581, \dodoi{10.1051/0004-6361:20066283}

\bibitem[{{Heard} \& {Gaskell}(2023)}]{2023MNRAS.518..418H}
{Heard}, C. Z.~P., \& {Gaskell}, C.~M. 2023, \mnras, 518, 418,
  \dodoi{10.1093/mnras/stac2220}

\bibitem[{{Hinshaw} {et~al.}(2013){Hinshaw}, {Larson}, {Komatsu}, {Spergel},
  {Bennett}, {Dunkley}, {Nolta}, {Halpern}, {Hill}, {Odegard}, {Page}, {Smith},
  {Weiland}, {Gold}, {Jarosik}, {Kogut}, {Limon}, {Meyer}, {Tucker}, {Wollack},
  \& {Wright}}]{2013ApJS..208...19H}
{Hinshaw}, G., {Larson}, D., {Komatsu}, E., {et~al.} 2013, \apjs, 208, 19,
  \dodoi{10.1088/0067-0049/208/2/19}

\bibitem[{{Hoormann} {et~al.}(2019){Hoormann}, {Martini}, {Davis}, {King},
  {Lidman}, {Mudd}, {Sharp}, {Sommer}, {Tucker}, {Yu}, {Allam}, {Asorey},
  {Avila}, {Banerji}, {Brooks}, {Buckley-Geer}, {Burke}, {Calcino}, {Carnero
  Rosell}, {Carollo}, {Carrasco Kind}, {Carretero}, {Castander}, {Childress},
  {De Vicente}, {Desai}, {Diehl}, {Doel}, {Flaugher}, {Fosalba}, {Frieman},
  {Garc{\'\i}a-Bellido}, {Gerdes}, {Gruen}, {Gutierrez}, {Hartley}, {Hinton},
  {Hollowood}, {Honscheid}, {Hoyle}, {James}, {Krause}, {Kuehn}, {Kuropatkin},
  {Lewis}, {Lima}, {Macaulay}, {Maia}, {Menanteau}, {Miller}, {Miquel},
  {M{\"o}ller}, {Plazas}, {Romer}, {Roodman}, {Sanchez}, {Scarpine},
  {Schubnell}, {Serrano}, {Sevilla-Noarbe}, {Smith}, {Smith}, {Soares-Santos},
  {Sobreira}, {Suchyta}, {Swann}, {Swanson}, {Tarle}, {Uddin}, \& {DES
  Collaboration}}]{2019MNRAS.487.3650H}
{Hoormann}, J.~K., {Martini}, P., {Davis}, T.~M., {et~al.} 2019, \mnras, 487,
  3650, \dodoi{10.1093/mnras/stz1539}

\bibitem[{{Hunter}(2007)}]{hunter2007}
{Hunter}, J.~D. 2007, Computing in Science and Engineering, 9, 90,
  \dodoi{10.1109/MCSE.2007.55}

\bibitem[{{Jaiswal} {et~al.}(2023){Jaiswal}, {Prince}, {Panda}, \&
  {Czerny}}]{2023A&A...670A.147J}
{Jaiswal}, V.~K., {Prince}, R., {Panda}, S., \& {Czerny}, B. 2023, \aap, 670,
  A147, \dodoi{10.1051/0004-6361/202244352}

\bibitem[{{Kammoun} {et~al.}(2021{\natexlab{a}}){Kammoun}, {Dov{\v{c}}iak},
  {Papadakis}, {Caballero-Garc{\'\i}a}, \& {Karas}}]{2021ApJ...907...20K}
{Kammoun}, E.~S., {Dov{\v{c}}iak}, M., {Papadakis}, I.~E.,
  {Caballero-Garc{\'\i}a}, M.~D., \& {Karas}, V. 2021{\natexlab{a}}, \apj, 907,
  20, \dodoi{10.3847/1538-4357/abcb93}

\bibitem[{{Kammoun} {et~al.}(2021{\natexlab{b}}){Kammoun}, {Papadakis}, \&
  {Dov{\v{c}}iak}}]{2021MNRAS.503.4163K}
{Kammoun}, E.~S., {Papadakis}, I.~E., \& {Dov{\v{c}}iak}, M.
  2021{\natexlab{b}}, \mnras, 503, 4163, \dodoi{10.1093/mnras/stab725}

\bibitem[{{Kaspi} {et~al.}(2021){Kaspi}, {Brandt}, {Maoz}, {Netzer},
  {Schneider}, {Shemmer}, \& {Grier}}]{2021ApJ...915..129K}
{Kaspi}, S., {Brandt}, W.~N., {Maoz}, D., {et~al.} 2021, \apj, 915, 129,
  \dodoi{10.3847/1538-4357/ac00aa}

\bibitem[{{Kaspi} {et~al.}(2000){Kaspi}, {Smith}, {Netzer}, {Maoz}, {Jannuzi},
  \& {Giveon}}]{2000ApJ...533..631K}
{Kaspi}, S., {Smith}, P.~S., {Netzer}, H., {et~al.} 2000, \apj, 533, 631,
  \dodoi{10.1086/308704}

\bibitem[{{Kinney} {et~al.}(1996){Kinney}, {Calzetti}, {Bohlin}, {McQuade},
  {Storchi-Bergmann}, \& {Schmitt}}]{1996ApJ...467...38K}
{Kinney}, A.~L., {Calzetti}, D., {Bohlin}, R.~C., {et~al.} 1996, \apj, 467, 38,
  \dodoi{10.1086/177583}

\bibitem[{{Koratkar} \& {Gaskell}(1991)}]{1991ApJ...370L..61K}
{Koratkar}, A.~P., \& {Gaskell}, C.~M. 1991, \apjl, 370, L61,
  \dodoi{10.1086/185977}

\bibitem[{{Korista} \& {Goad}(2001)}]{2001ApJ...553..695K}
{Korista}, K.~T., \& {Goad}, M.~R. 2001, \apj, 553, 695, \dodoi{10.1086/320964}

\bibitem[{{Korista} \& {Goad}(2019)}]{2019MNRAS.489.5284K}
---. 2019, \mnras, 489, 5284, \dodoi{10.1093/mnras/stz2330}

\bibitem[{{Lasota}(2016)}]{2016ASSL..440....1L}
{Lasota}, J.-P. 2016, in Astrophysics and Space Science Library, Vol. 440,
  Astrophysics of Black Holes: From Fundamental Aspects to Latest Developments,
  ed. C.~{Bambi}, 1, \dodoi{10.1007/978-3-662-52859-4_1}

\bibitem[{{Lawther} {et~al.}(2018){Lawther}, {Goad}, {Korista}, {Ulrich}, \&
  {Vestergaard}}]{2018MNRAS.481..533L}
{Lawther}, D., {Goad}, M.~R., {Korista}, K.~T., {Ulrich}, O., \& {Vestergaard},
  M. 2018, \mnras, 481, 533, \dodoi{10.1093/mnras/sty2242}

\bibitem[{{Lira} {et~al.}(2018){Lira}, {Kaspi}, {Netzer}, {Botti}, {Morrell},
  {Mej{\'\i}a-Restrepo}, {S{\'a}nchez-S{\'a}ez}, {Mart{\'\i}nez-Palomera}, \&
  {L{\'o}pez}}]{2018ApJ...865...56L}
{Lira}, P., {Kaspi}, S., {Netzer}, H., {et~al.} 2018, \apj, 865, 56,
  \dodoi{10.3847/1538-4357/aada45}

\bibitem[{{Lobban} {et~al.}(2018){Lobban}, {Porquet}, {Reeves}, {Markowitz},
  {Nardini}, \& {Grosso}}]{2018MNRAS.474.3237L}
{Lobban}, A.~P., {Porquet}, D., {Reeves}, J.~N., {et~al.} 2018, \mnras, 474,
  3237, \dodoi{10.1093/mnras/stx2889}

\bibitem[{{Loiacono} {et~al.}(2024){Loiacono}, {Decarli}, {Mignoli}, {Farina},
  {Ba{\~n}ados}, {Bosman}, {Eilers}, {Schindler}, {Strauss}, {Vestergaard},
  {Wang}, {Blecha}, {Carilli}, {Comastri}, {Connor}, {Costa}, {Dotti}, {Fan},
  {Gilli}, {Jun}, {Liu}, {Lupi}, {Marshall}, {Mazzucchelli}, {Meyer},
  {Neeleman}, {Overzier}, {Pensabene}, {Riechers}, {Trakhtenbrot}, {Trebitsch},
  {Venemans}, {Walter}, \& {Yang}}]{2024arXiv240213319L}
{Loiacono}, F., {Decarli}, R., {Mignoli}, M., {et~al.} 2024, arXiv e-prints,
  arXiv:2402.13319, \dodoi{10.48550/arXiv.2402.13319}

\bibitem[{{Lynden-Bell}(1969)}]{1969Natur.223..690L}
{Lynden-Bell}, D. 1969, \nat, 223, 690, \dodoi{10.1038/223690a0}

\bibitem[{{Marziani} {et~al.}(2016){Marziani}, {Mart{\'\i}nez Carballo},
  {Sulentic}, {Del Olmo}, {Stirpe}, \& {Dultzin}}]{2016Ap&SS.361...29M}
{Marziani}, P., {Mart{\'\i}nez Carballo}, M.~A., {Sulentic}, J.~W., {et~al.}
  2016, \apss, 361, 29, \dodoi{10.1007/s10509-015-2611-1}

\bibitem[{{Marziani} {et~al.}(2024){Marziani}, {Floris}, {Deconto-Machado},
  {Panda}, {Sniegowska}, {Garnica}, {Dultzin}, {D'Onofrio}, {Del Olmo}, {Bon},
  \& {Bon}}]{2024Physi...6..216M}
{Marziani}, P., {Floris}, A., {Deconto-Machado}, A., {et~al.} 2024, Physics, 6,
  216, \dodoi{10.3390/physics6010016}

\bibitem[{{Maza} {et~al.}(1993){Maza}, {Ruiz}, {Gonzalez}, {Wischnjewsky}, \&
  {Antezana}}]{1993RMxAA..25...51M}
{Maza}, J., {Ruiz}, M.~T., {Gonzalez}, L.~E., {Wischnjewsky}, M., \&
  {Antezana}, R. 1993, \rmxaa, 25, 51

\bibitem[{{Maza} {et~al.}(1995){Maza}, {Wischnjewsky}, {Antezana}, \&
  {Gonz{\'a}lez}}]{1995RMxAA..31..119M}
{Maza}, J., {Wischnjewsky}, M., {Antezana}, R., \& {Gonz{\'a}lez}, L.~E. 1995,
  \rmxaa, 31, 119

\bibitem[{{McLure} \& {Dunlop}(2004)}]{2004MNRAS.352.1390M}
{McLure}, R.~J., \& {Dunlop}, J.~S. 2004, \mnras, 352, 1390,
  \dodoi{10.1111/j.1365-2966.2004.08034.x}

\bibitem[{{Mej{\'\i}a-Restrepo} {et~al.}(2018){Mej{\'\i}a-Restrepo}, {Lira},
  {Netzer}, {Trakhtenbrot}, \& {Capellupo}}]{2018NatAs...2...63M}
{Mej{\'\i}a-Restrepo}, J.~E., {Lira}, P., {Netzer}, H., {Trakhtenbrot}, B., \&
  {Capellupo}, D.~M. 2018, Nature Astronomy, 2, 63,
  \dodoi{10.1038/s41550-017-0305-z}

\bibitem[{{Morgan} {et~al.}(2012){Morgan}, {Hainline}, {Chen}, {Tewes},
  {Kochanek}, {Dai}, {Kozlowski}, {Blackburne}, {Mosquera}, {Chartas},
  {Courbin}, \& {Meylan}}]{2012ApJ...756...52M}
{Morgan}, C.~W., {Hainline}, L.~J., {Chen}, B., {et~al.} 2012, \apj, 756, 52,
  \dodoi{10.1088/0004-637X/756/1/52}

\bibitem[{{Mosquera} {et~al.}(2013){Mosquera}, {Kochanek}, {Chen}, {Dai},
  {Blackburne}, \& {Chartas}}]{2013ApJ...769...53M}
{Mosquera}, A.~M., {Kochanek}, C.~S., {Chen}, B., {et~al.} 2013, \apj, 769, 53,
  \dodoi{10.1088/0004-637X/769/1/53}

\bibitem[{{Narayan} \& {Yi}(1994)}]{1994ApJ...428L..13N}
{Narayan}, R., \& {Yi}, I. 1994, \apjl, 428, L13, \dodoi{10.1086/187381}

\bibitem[{{Negrete} {et~al.}(2014){Negrete}, {Dultzin}, {Marziani}, \&
  {Sulentic}}]{2014AdSpR..54.1355N}
{Negrete}, C.~A., {Dultzin}, D., {Marziani}, P., \& {Sulentic}, J.~W. 2014,
  Advances in Space Research, 54, 1355, \dodoi{10.1016/j.asr.2013.11.037}

\bibitem[{{Netzer}(1990)}]{1990agn..conf...57N}
{Netzer}, H. 1990, in Active Galactic Nuclei, ed. R.~D. {Blandford},
  H.~{Netzer}, L.~{Woltjer}, T.~J.~L. {Courvoisier}, \& M.~{Mayor}, 57--160

\bibitem[{{Netzer}(2022)}]{2022MNRAS.509.2637N}
{Netzer}, H. 2022, \mnras, 509, 2637, \dodoi{10.1093/mnras/stab3133}

\bibitem[{{Novikov} \& {Thorne}(1973)}]{1973blho.conf..343N}
{Novikov}, I.~D., \& {Thorne}, K.~S. 1973, in Black Holes (Les Astres Occlus),
  343--450

\bibitem[{{Panda}(2021)}]{2021A&A...650A.154P}
{Panda}, S. 2021, \aap, 650, A154, \dodoi{10.1051/0004-6361/202140393}

\bibitem[{{Panda} {et~al.}(2018){Panda}, {Czerny}, {Adhikari}, {Hryniewicz},
  {Wildy}, {Kuraszkiewicz}, \& {{\'S}niegowska}}]{2018ApJ...866..115P}
{Panda}, S., {Czerny}, B., {Adhikari}, T.~P., {et~al.} 2018, \apj, 866, 115,
  \dodoi{10.3847/1538-4357/aae209}

\bibitem[{{Panda} {et~al.}(2019{\natexlab{a}}){Panda}, {Czerny}, {Done}, \&
  {Kubota}}]{2019ApJ...875..133P}
{Panda}, S., {Czerny}, B., {Done}, C., \& {Kubota}, A. 2019{\natexlab{a}},
  \apj, 875, 133, \dodoi{10.3847/1538-4357/ab11cb}

\bibitem[{{Panda} {et~al.}(2019{\natexlab{b}}){Panda}, {Mart{\'\i}nez-Aldama},
  \& {Zaja{\v{c}}ek}}]{2019FrASS...6...75P}
{Panda}, S., {Mart{\'\i}nez-Aldama}, M.~L., \& {Zaja{\v{c}}ek}, M.
  2019{\natexlab{b}}, Frontiers in Astronomy and Space Sciences, 6, 75,
  \dodoi{10.3389/fspas.2019.00075}

\bibitem[{{Panda} \& {Marziani}(2023{\natexlab{a}})}]{2023FrASS..1030103P}
{Panda}, S., \& {Marziani}, P. 2023{\natexlab{a}}, Frontiers in Astronomy and
  Space Sciences, 10, 1130103, \dodoi{10.3389/fspas.2023.1130103}

\bibitem[{{Panda} \& {Marziani}(2023{\natexlab{b}})}]{2023BoSAB..34..241P}
---. 2023{\natexlab{b}}, Boletim da Sociedade Astronomica Brasileira, 34, 241,
  \dodoi{10.48550/arXiv.2308.05830}

\bibitem[{{Panda} {et~al.}(2023){Panda}, {Marziani}, {Czerny},
  {Rodr{\'\i}guez-Ardila}, \& {Pozo Nu{\~n}ez}}]{2023Univ....9..492P}
{Panda}, S., {Marziani}, P., {Czerny}, B., {Rodr{\'\i}guez-Ardila}, A., \&
  {Pozo Nu{\~n}ez}, F. 2023, Universe, 9, 492, \dodoi{10.3390/universe9120492}

\bibitem[{{Pandey} {et~al.}(2023){Pandey}, {Czerny}, {Panda}, {Prince},
  {Jaiswal}, {Martinez-Aldama}, {Zaja{\v{c}}ek}, \&
  {{\'S}niegowska}}]{2023A&A...680A.102P}
{Pandey}, A., {Czerny}, B., {Panda}, S., {et~al.} 2023, \aap, 680, A102,
  \dodoi{10.1051/0004-6361/202347819}

\bibitem[{{Pandey} {et~al.}(2024){Pandey}, {Mart{\'\i}nez-Aldama}, {Czerny},
  {Panda}, \& {Zaja{\v{c}}ek}}]{2024arXiv240118052P}
{Pandey}, A., {Mart{\'\i}nez-Aldama}, M.~L., {Czerny}, B., {Panda}, S., \&
  {Zaja{\v{c}}ek}, M. 2024, arXiv e-prints, arXiv:2401.18052,
  \dodoi{10.48550/arXiv.2401.18052}

\bibitem[{{Papadakis} {et~al.}(2022){Papadakis}, {Dov{\v{c}}iak}, \&
  {Kammoun}}]{2022A&A...666A..11P}
{Papadakis}, I.~E., {Dov{\v{c}}iak}, M., \& {Kammoun}, E.~S. 2022, \aap, 666,
  A11, \dodoi{10.1051/0004-6361/202142962}

\bibitem[{{Pooley} {et~al.}(2007){Pooley}, {Blackburne}, {Rappaport}, \&
  {Schechter}}]{2007ApJ...661...19P}
{Pooley}, D., {Blackburne}, J.~A., {Rappaport}, S., \& {Schechter}, P.~L. 2007,
  \apj, 661, 19, \dodoi{10.1086/512115}

\bibitem[{{Popovi{\'c}} {et~al.}(2019){Popovi{\'c}},
  {Kova{\v{c}}evi{\'c}-Doj{\v{c}}inovi{\'c}}, \&
  {Mar{\v{c}}eta-Mandi{\'c}}}]{2019MNRAS.484.3180P}
{Popovi{\'c}}, L.~{\v{C}}., {Kova{\v{c}}evi{\'c}-Doj{\v{c}}inovi{\'c}}, J., \&
  {Mar{\v{c}}eta-Mandi{\'c}}, S. 2019, \mnras, 484, 3180,
  \dodoi{10.1093/mnras/stz157}

\bibitem[{{Pozo Nu{\~n}ez} {et~al.}(2023{\natexlab{a}}){Pozo Nu{\~n}ez},
  {Bruckmann}, {Deesamutara}, {Czerny}, {Panda}, {Lobban}, {Pietrzy{\'n}ski},
  \& {Polsterer}}]{2023MNRAS.522.2002P}
{Pozo Nu{\~n}ez}, F., {Bruckmann}, C., {Deesamutara}, S., {et~al.}
  2023{\natexlab{a}}, \mnras, 522, 2002, \dodoi{10.1093/mnras/stad286}

\bibitem[{{Pozo Nu{\~n}ez} {et~al.}(2017){Pozo Nu{\~n}ez}, {Chelouche},
  {Kaspi}, \& {Niv}}]{2017PASP..129i4101P}
{Pozo Nu{\~n}ez}, F., {Chelouche}, D., {Kaspi}, S., \& {Niv}, S. 2017, \pasp,
  129, 094101, \dodoi{10.1088/1538-3873/aa7a55}

\bibitem[{{Pozo Nu{\~n}ez} {et~al.}(2023{\natexlab{b}}){Pozo Nu{\~n}ez},
  {Gianniotis}, \& {Polsterer}}]{2023A&A...674A..83P}
{Pozo Nu{\~n}ez}, F., {Gianniotis}, N., \& {Polsterer}, K.~L.
  2023{\natexlab{b}}, \aap, 674, A83, \dodoi{10.1051/0004-6361/202345932}

\bibitem[{{Pozo Nu{\~n}ez} {et~al.}(2019){Pozo Nu{\~n}ez}, {Gianniotis},
  {Blex}, {Lisow}, {Chini}, {Polsterer}, {Pott}, {Esser}, \&
  {Pietrzy{\'n}ski}}]{2019MNRAS.490.3936P}
{Pozo Nu{\~n}ez}, F., {Gianniotis}, N., {Blex}, J., {et~al.} 2019, \mnras, 490,
  3936, \dodoi{10.1093/mnras/stz2830}

\bibitem[{{Prieto} {et~al.}(2022){Prieto}, {Rodr{\'\i}guez-Ardila}, {Panda}, \&
  {Marinello}}]{2022MNRAS.510.1010P}
{Prieto}, A., {Rodr{\'\i}guez-Ardila}, A., {Panda}, S., \& {Marinello}, M.
  2022, \mnras, 510, 1010, \dodoi{10.1093/mnras/stab3414}

\bibitem[{{Pringle}(1981)}]{1981ARA&A..19..137P}
{Pringle}, J.~E. 1981, \araa, 19, 137,
  \dodoi{10.1146/annurev.aa.19.090181.001033}

\bibitem[{{Pringle} \& {Rees}(1972)}]{1972A&A....21....1P}
{Pringle}, J.~E., \& {Rees}, M.~J. 1972, \aap, 21, 1

\bibitem[{{Reynolds}(2019)}]{2019NatAs...3...41R}
{Reynolds}, C.~S. 2019, Nature Astronomy, 3, 41,
  \dodoi{10.1038/s41550-018-0665-z}

\bibitem[{{Sergeev} {et~al.}(2005){Sergeev}, {Doroshenko}, {Golubinskiy},
  {Merkulova}, \& {Sergeeva}}]{2005ApJ...622..129S}
{Sergeev}, S.~G., {Doroshenko}, V.~T., {Golubinskiy}, Y.~V., {Merkulova},
  N.~I., \& {Sergeeva}, E.~A. 2005, \apj, 622, 129, \dodoi{10.1086/427820}

\bibitem[{{Shakura} \& {Sunyaev}(1973)}]{1973A&A....24..337S}
{Shakura}, N.~I., \& {Sunyaev}, R.~A. 1973, \aap, 24, 337

\bibitem[{{Shankar} {et~al.}(2009){Shankar}, {Weinberg}, \&
  {Miralda-Escud{\'e}}}]{2009ApJ...690...20S}
{Shankar}, F., {Weinberg}, D.~H., \& {Miralda-Escud{\'e}}, J. 2009, \apj, 690,
  20, \dodoi{10.1088/0004-637X/690/1/20}

\bibitem[{{Shappee} {et~al.}(2014){Shappee}, {Prieto}, {Grupe}, {Kochanek},
  {Stanek}, {De Rosa}, {Mathur}, {Zu}, {Peterson}, {Pogge}, {Komossa}, {Im},
  {Jencson}, {Holoien}, {Basu}, {Beacom}, {Szczygie{\l}}, {Brimacombe},
  {Adams}, {Campillay}, {Choi}, {Contreras}, {Dietrich}, {Dubberley},
  {Elphick}, {Foale}, {Giustini}, {Gonzalez}, {Hawkins}, {Howell}, {Hsiao},
  {Koss}, {Leighly}, {Morrell}, {Mudd}, {Mullins}, {Nugent}, {Parrent},
  {Phillips}, {Pojmanski}, {Rosing}, {Ross}, {Sand}, {Terndrup}, {Valenti},
  {Walker}, \& {Yoon}}]{2014ApJ...788...48S}
{Shappee}, B.~J., {Prieto}, J.~L., {Grupe}, D., {et~al.} 2014, \apj, 788, 48,
  \dodoi{10.1088/0004-637X/788/1/48}

\bibitem[{{Shen} {et~al.}(2023){Shen}, {Grier}, {Horne}, {Stone}, {Li}, {Yang},
  {Homayouni}, {Trump}, {Anderson}, {Brandt}, {Hall}, {Ho}, {Jiang},
  {Petitjean}, {Schneider}, {Tao}, {Donnan}, {AlSayyad}, {Bershady}, {Blanton},
  {Bizyaev}, {Bundy}, {Chen}, {Davis}, {Dawson}, {Fan}, {Greene}, {Groller},
  {Guo}, {Ibarra-Medel}, {Jiang}, {Keenan}, {Kollmeier}, {Lejoly}, {Li}, {de la
  Macorra}, {Moe}, {Nie}, {Rossi}, {Smith}, {Tee}, {Weijmans}, {Xu}, {Yue},
  {Zhou}, {Zhou}, \& {Zou}}]{2023arXiv230501014S}
{Shen}, Y., {Grier}, C.~J., {Horne}, K., {et~al.} 2023, arXiv e-prints,
  arXiv:2305.01014, \dodoi{10.48550/arXiv.2305.01014}

\bibitem[{{Sulentic} {et~al.}(2017){Sulentic}, {del Olmo}, {Marziani},
  {Mart{\'\i}nez-Carballo}, {D'Onofrio}, {Dultzin}, {Perea},
  {Mart{\'\i}nez-Aldama}, {Negrete}, {Stirpe}, \&
  {Zamfir}}]{2017A&A...608A.122S}
{Sulentic}, J.~W., {del Olmo}, A., {Marziani}, P., {et~al.} 2017, \aap, 608,
  A122, \dodoi{10.1051/0004-6361/201630309}

\bibitem[{{Timmer} \& {Koenig}(1995)}]{1995A&A...300..707T}
{Timmer}, J., \& {Koenig}, M. 1995, \aap, 300, 707

\bibitem[{{van der Walt} {et~al.}(2011){van der Walt}, {Colbert}, \&
  {Varoquaux}}]{van2011}
{van der Walt}, S., {Colbert}, S.~C., \& {Varoquaux}, G. 2011, Computing in
  Science and Engineering, 13, 22, \dodoi{10.1109/MCSE.2011.37}

\bibitem[{{Vaughan} {et~al.}(2003){Vaughan}, {Edelson}, {Warwick}, \&
  {Uttley}}]{2003MNRAS.345.1271V}
{Vaughan}, S., {Edelson}, R., {Warwick}, R.~S., \& {Uttley}, P. 2003, \mnras,
  345, 1271, \dodoi{10.1046/j.1365-2966.2003.07042.x}

\bibitem[{{Vietri} {et~al.}(2020){Vietri}, {Mainieri}, {Kakkad}, {Netzer},
  {Perna}, {Circosta}, {Harrison}, {Zappacosta}, {Husemann}, {Padovani},
  {Bischetti}, {Bongiorno}, {Brusa}, {Carniani}, {Cicone}, {Comastri},
  {Cresci}, {Feruglio}, {Fiore}, {Lanzuisi}, {Mannucci}, {Marconi},
  {Piconcelli}, {Puglisi}, {Salvato}, {Schramm}, {Schulze}, {Scholtz},
  {Vignali}, \& {Zamorani}}]{2020A&A...644A.175V}
{Vietri}, G., {Mainieri}, V., {Kakkad}, D., {et~al.} 2020, \aap, 644, A175,
  \dodoi{10.1051/0004-6361/202039136}

\bibitem[{{Virtanen} {et~al.}(2020){Virtanen}, {Gommers}, {Oliphant},
  {Haberland}, {Reddy}, {Cournapeau}, {Burovski}, {Peterson}, {Weckesser},
  {Bright}, {van der Walt}, {Brett}, {Wilson}, {Millman}, {Mayorov}, {Nelson},
  {Jones}, {Kern}, {Larson}, {Carey}, {Polat}, {Feng}, {Moore}, {VanderPlas},
  {Laxalde}, {Perktold}, {Cimrman}, {Henriksen}, {Quintero}, {Harris},
  {Archibald}, {Ribeiro}, {Pedregosa}, {van Mulbregt}, \& {SciPy 1. 0
  Contributors}}]{virtanen2020}
{Virtanen}, P., {Gommers}, R., {Oliphant}, T.~E., {et~al.} 2020, Nature
  Methods, 17, 261, \dodoi{10.1038/s41592-019-0686-2}

\bibitem[{{Wandel} {et~al.}(1999){Wandel}, {Peterson}, \&
  {Malkan}}]{1999ApJ...526..579W}
{Wandel}, A., {Peterson}, B.~M., \& {Malkan}, M.~A. 1999, \apj, 526, 579,
  \dodoi{10.1086/308017}

\bibitem[{{Wang} {et~al.}(2023){Wang}, {Guo}, \& {Woo}}]{2023ApJ...948L..23W}
{Wang}, S., {Guo}, H., \& {Woo}, J.-H. 2023, \apjl, 948, L23,
  \dodoi{10.3847/2041-8213/accf96}

\bibitem[{{Wang} {et~al.}(2019){Wang}, {Shen}, {Jiang}, {Horne}, {Brandt},
  {Grier}, {Ho}, {Homayouni}, {I-Hsiu Li}, {Schneider}, \&
  {Trump}}]{2019ApJ...882....4W}
{Wang}, S., {Shen}, Y., {Jiang}, L., {et~al.} 2019, \apj, 882, 4,
  \dodoi{10.3847/1538-4357/ab322b}

\bibitem[{{Yuan} \& {Narayan}(2014)}]{2014ARA&A..52..529Y}
{Yuan}, F., \& {Narayan}, R. 2014, \araa, 52, 529,
  \dodoi{10.1146/annurev-astro-082812-141003}

\bibitem[{{Zaja{\v{c}}ek} {et~al.}(2023){Zaja{\v{c}}ek}, {Panda}, {Pandey},
  {Prince}, {Rodr{\'\i}guez-Ardila}, {Jaiswal}, {Czerny}, {Hryniewicz},
  {Urbanowicz}, {Trzcionkowski}, {{\'S}niegowska}, {Fa{\l}kowska},
  {Mart{\'\i}nez-Aldama}, \& {Werner}}]{2023arXiv231003544Z}
{Zaja{\v{c}}ek}, M., {Panda}, S., {Pandey}, A., {et~al.} 2023, arXiv e-prints,
  arXiv:2310.03544, \dodoi{10.48550/arXiv.2310.03544}

\end{thebibliography}
\bibliographystyle{aasjournal}



\end{document}